\documentclass[journal,twoside]{IEEEtran}

\usepackage{graphicx}
\usepackage [noadjust]{cite} 
\usepackage{bm}  
\usepackage{amsmath}
\usepackage{amssymb}
\usepackage{enumerate}
\usepackage{cases}
\usepackage{epstopdf}
\usepackage[percent]{overpic}

\usepackage{soul} 

\usepackage{tikz}
\usepackage{pgfplots}
\usetikzlibrary{spy,calc}

 \newcommand{\A}{\mathcal{A}}
 \newcommand{\B}{\mathcal{B}}
 \newcommand{\s}{\mathcal{S}}
 \newcommand{\M}{\mathcal{M}}
 
 \newcommand{\argmax}{\operatornamewithlimits{argmax}}
 \newcommand{\argmin}{\operatornamewithlimits{argmin}}

 \newcommand{\pimax}{p_i^{\mathrm{max}}}
 \newcommand{\pbold}{\mathbf{p}}
 \newcommand{\pboldzero}{\mathring{\mathbf{p}}}
 \newcommand{\pzero}{\mathring{p}}

 \newcommand{\gammai}{\gamma_i}

 \newcommand{\Phibold}{\boldsymbol{\Phi}}

 \newcommand{\gammaihat}{\mathit{\widehat{\gamma}}_i}
 \newcommand{\gammahatbold}{\widehat{\mbox{\boldmath{$\gamma$}}}}
 \newcommand{\gammahat}{\mathit{\widehat{\gamma}}}

 \newcommand{\Fbold}{\mathbf{F}}

 \newcommand{\startalg}[1]{\newcounter{algcounter} 	
 							\stepcounter{algcounter} #1
 							\newcounter{algblockspace}
 							\setcounter{algblockspace}{0}}
 \newcommand{\startalgg}[1]{\setcounter{algcounter}{1} #1
				 			\setcounter{algblockspace}{0}}
 \newcommand{\algstartblock}{\addtocounter{algblockspace}{16}}
 \newcommand{\algendblock}{\addtocounter{algblockspace}{-16}}
 
 \newcommand{\alg}[1]{\begin{enumerate}[\arabic{algcounter}.\hspace{\value{algblockspace}pt}]
  				\item #1  \end{enumerate} \stepcounter{algcounter}}

\newcommand{\gammabold}{\boldsymbol{\gamma}}
\newcommand{\Gammabold}{\boldsymbol{\Gamma}}

 \newcommand{\hup}[2]{h_{#1 #2}}
 
 \newcommand{\Nup}[1]{N_{#1}}
 
 \newcommand{\pup}{p}

\newcommand{\pmax}{p^{\mathrm{max}}}

\newcommand{\CIbold}{\mathbf{I}^{s \rightarrow p}}
\newcommand{\CIzero}{{\mathring{I}^{s \rightarrow p}}}
\newcommand{\CIzerobold}{{\mathring{\mathbf{I}}^{s \rightarrow p}}}
\newtheorem {theorem}{Theorem}
\newtheorem {corollary}{Corollary}
\newtheorem {lemma}{Lemma}

\newtheorem {remark}{Remark}
\newtheorem {definition}{Definition}

\begin{document}

\raggedbottom

\title{On Characterization of Feasible Interference Regions in Cognitive Radio Networks}

\author{Mehdi~Monemi, Mehdi~Rasti, and Ekram Hossain
\thanks{M. Monemi is with the Dept. of Electrical and Computer Engineering,  Neyriz Branch, Islamic Azad University,  Neyriz, Iran (email: m\_monemi@shirazu.ac.ir). M. Rasti is with the Dept. of Computer Engineering and Information Technology, Amirkabir University of Technology, Tehran, Iran (email: rasti@aut.ac.ir).
E. Hossain is with the Dept. of Electrical and Computer Engineering, University of Manitoba, Canada (email: Ekram.Hossain@umanitoba.ca).}
}

\maketitle

\begin{abstract}
	In an underlay cognitive radio network (CRN), in order to guarantee that all primary users (PUs) achieve their target-signal-to-interference-plus-noise ratios (target-SINRs), the interference caused by all secondary users (SUs) to the primary receiving-points should be controlled. To do so, the feasible cognitive interference region (FCIR), i.e., the region for allowable values of interference at all of the primary receiving-points, which guarantee the protection of the PUs, needs to be formally characterized.
	In the state-of-the-art interference management schemes for underlay CRNs, it is considered that all PUs are protected if the cognitive interference for each primary receiving-point is lower than a  maximum threshold, the so called interference temperature limit (ITL) for the corresponding receiving-point. This is assumed to be fixed and independent of ITL values for other primary receiving-points, which corresponds to a box-like FCIR. In this paper, we  characterize the FCIR for {\em uplink} transmissions in cellular CRNs and for direct transmissions in ad-hoc CRNs. We  show that  the FCIR is in fact a polyhedron (i.e.,  the maximum feasible cognitive interference threshold for each primary receiving-point is not a constant, and it depends on that for the other primary receiving-points). Therefore, in practical interference management algorithms, it is not proper to consider a constant and independent ITL value for each of the primary receiving-points. This finding would significantly affect the design of practical interference management schemes for CRNs. To demonstrate  this, based on the characterized FCIR, we  propose two power control algorithms to find the maximum number of admitted SUs and the maximum aggregate throughput of the SUs in infeasible and feasible CRNs, respectively. For two distinct objectives, our proposed interference management schemes  outperform the existing ones. 
	The numerical results also demonstrate how the assumption of fixed ITL values leads to poor performance measures in CRNs.
\end{abstract}


\thispagestyle{empty}

\section{Introduction}

	In order to improve the utilization of radio spectrum, cognitive radio (CR) has been the focus of many of recent research studies. In a cognitive radio network (CRN), unlicensed secondary users (SUs) coexist with a primary radio network (PRN) serving licensed primary users (PUs). The PRN can dynamically share the spectrum with the  SUs so that the admitted SUs achieve their minimum acceptable quality-of-service (QoS), and at the same time, all the PUs are protected. That is, the  SUs do not violate the QoS requirements of the PUs.
	
	Cognitive radio systems can use either spectrum overlay or spectrum underlay strategy for channel access. In case of spectrum overlay, the channels which are unused by the PUs are detected by the SUs through spectrum sensing mechanisms and are used by the SUs. In the spectrum underlay scenario,  the entire frequency spectrum is shared by all of the PUs and SUs. In this case, since admission of any  SU causes interference to all of the PUs' receiving points, the interference caused by the SUs must be controlled through power control strategies in a way that all PUs are protected (e.g., all PUs achieve their target signal-to-interference-plus-noise ratios [SINRs]).

	There are two major interference management problems for underlay CRNs which have been considered  in the literature. In an infeasible system, where all PUs and SUs may not be simultaneously supported with their target-SINRs, it is generally desired to devise a joint power and admission control (JPAC) algorithm that may protect all PUs together with maximum number of admitted SUs.
	In a feasible system where all SUs may attain SINRs higher than their minimum acceptable target-SINRs, it is generally desirable to assign SUs as high SINR values possible as to maximize some QoS measure (e.g., aggregate throughput of the SUs) subject to the constraint of PUs' protection.
	In these two kinds of interference management problems, the corresponding objective function for CRN is optimized subject to the \textit{protection constraints for the PUs}. 
	In \cite{distributed_JPAC_antenna_arrays}, a distributed algorithm is introduced to minimize the total transmit power of primary and secondary links by using antenna arrays. In \cite{JPAC_adhoc_convex_relaxation_single_PU}, a distributed JPAC algorithm is proposed for ad-hoc networks by convex relaxation of the non-convex problem to obtain the maximum number of supported SUs for a CRN coexisting with a single PU. In \cite{7, JPAC_SSA1,ISMIRA, LGRA,monemi_ESRPA}, several centralized JPAC algorithms are proposed to obtain suboptimal solutions to the problem of maximum number of supported SUs. In \cite{throughput_in_mesh_networks,bad_ITL6_IET_throughput,GP_bad_ITL,badITL4,throughput_coupled_interference_game_bad_ITL,Robust_Max_Throughput_Wang_2015,throughput_distributed_game_bad_ITL,Beamforming_Cognitive_2015,Max_Throughput_Femto_Zhang_2015} several centralized and distributed algorithms are proposed to maximize the aggregate throughput of SUs in feasible cognitive radio networks. Subject to the protection of the high-prioritized macrocell users, several power control optimization problems are also investigated for the two-tier macrocell-femtocell networks in \cite{femto_tow_tier_ngo_1,femto_tow_tier_ngo_2,hierarchical_tow_tier_guruacharya,Max_Throughput_Femto_Zhang_2015}. The calculation of the capacity of cognitive radio networks by using distributed virtual antenna arrays (VAAs) and a multi-antenna BS is investigated in \cite{wang5}. In \cite{wang4}, the aggregate interference models for a CRN coexisting with a single primary transmitter-receiver link is investigated.

\begin{figure}
		\centering
		\includegraphics [width=200pt]{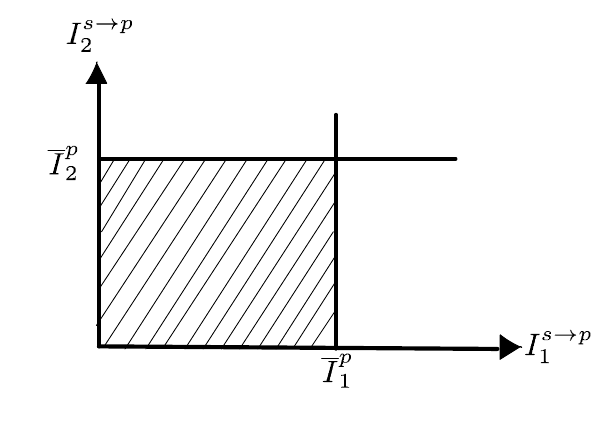}
		\vspace{-15pt}
		\caption{General concept of the FCIR in the state-of-the-art interference management algorithms based on the ITL values in a network consisting of two primary receivers. Here $\overline{I}_1^{p}$ and $\overline{I}_2^{p}$ are the so-called ITL values for the first and second primary receiving-points, respectively, and $I^{s\rightarrow p}_1$ and $I^{s\rightarrow p}_2$ are the interference imposed from all the SUs to the first and second primary receiving-points, respectively.} \vspace{-10pt}
		\label{fig:ex_TLFCIR}
	\end{figure}

	To deal with interference management problems, the {\em feasible cognitive interference region (FCIR)}, i.e., the region for allowable values of the cognitive interference (i.e., interference caused by the SUs) at any of the primary receiving-points which guarantee  that the \textit{protection constraints for the PUs} are satisfied, needs to be formally characterized. 
	In the state-of-the-art interference management schemes for underlay CRNs 
	(e.g.,  \cite{JPAC_adhoc_convex_relaxation_single_PU, distributed_JPAC_antenna_arrays, 7, JPAC_SSA1,ISMIRA, LGRA,monemi_ESRPA,throughput_in_mesh_networks,bad_ITL6_IET_throughput,GP_bad_ITL,badITL4,throughput_coupled_interference_game_bad_ITL,Robust_Max_Throughput_Wang_2015,throughput_distributed_game_bad_ITL,Beamforming_Cognitive_2015,Max_Throughput_Femto_Zhang_2015 }), it is assumed that all PUs are protected (the  protection constraints for the PUs are satisfied), if the cognitive interference for each primary receiving-point is lower than a  maximum threshold, which is called the interference temperature limit (ITL) for the corresponding receiving-point. This ITL is considered to be fixed and independent of ITL values for other primary receiving-points. 		
	In other words, the  PUs are assumed to be protected if the cognitive interference caused to each of the primary receiving-points is smaller than a constant ITL value for that receiving-point. This corresponds to a ``box-like" feasible region of the cognitive interference imposed at the primary receiving-points (Fig.~\ref{fig:ex_TLFCIR}). 

	In the above context, our contributions in this paper can be summarized as follows.
	\begin{itemize}
	\item
		We show that the box-like feasible region of cognitive interference caused to  the primary receiving-points considered in the literature is not equivalent to the SINR constraints (i.e., protection constraints) for the PUs. Specifically, we analytically define and obtain the FCIR and show that, as opposed to the  common intuition, the feasible cognitive interference region is a polyhedron. That is, the maximum feasible cognitive interference threshold for each primary receiving-point is not a constant, and it depends on the maximum feasible cognitive interference threshold for other primary receiving-points.
		Therefore, it is not proper to consider a constant and independent ITL value for each of the primary receiving-points. This discovery  would significantly affect the design of practical interference management schemes for CRNs. Since the ITL values for most of the existing interference management problems and their corresponding algorithms are given based on the box-like assumption, they need to be re-defined and re-designed, respectively, to achieve higher network performance. This can be performed based on the polyhedron  region characterized in this paper for the feasible cognitive interference caused by the SUs to the primary receiving-points. Besides, the characterized feasible region of the cognitive interference helps us to state and solve optimal resource allocation problems for the cognitive radios at the CRN level with minimal involvement of the primary radio network (PRN).
	\item
		To demonstrate how  the newly derived characterization of FCIR affects the system performance measures, we consider two cases of infeasible and feasible systems. For an infeasible system, we propose a novel low-complexity admission control algorithm that can be run by the CRN with minimal feedback from the PRN aiming at protecting all the PUs and supporting the maximum number of SUs. For a feasible system, to obtain the maximum aggregate throughput of SUs, we revise one of the existing algorithms by considering the constraint that the cognitive interference of all SUs imposed on the primary receiving-points lie within the characterized FCIR. 
	\end{itemize}
	
	The organization of the rest of the paper is as follows.  In Section \ref{sec:system_model}, the system model and background  are presented. In Section \ref{sec:charecterizing_FIR}, we formally characterize the feasible interference region for the primary receiving-points of the PRN. In Section \ref{sec:proposed_algorithm}, based on the characterized feasible interference region, we propose two power control algorithms for infeasible and feasible CRNs, respectively. Finally, numerical results  are presented in Section \ref{sec:numerical_results} before the paper is concluded in Section \ref{sec:conclusion}.
 
	
\section{System Model and Background}
	\label{sec:system_model}
	\subsection{System Model and Notations}
\begin{figure}
		\centering
		\setlength{\fboxrule}{0pt}
		\fbox{\begin{overpic}[width=250pt]{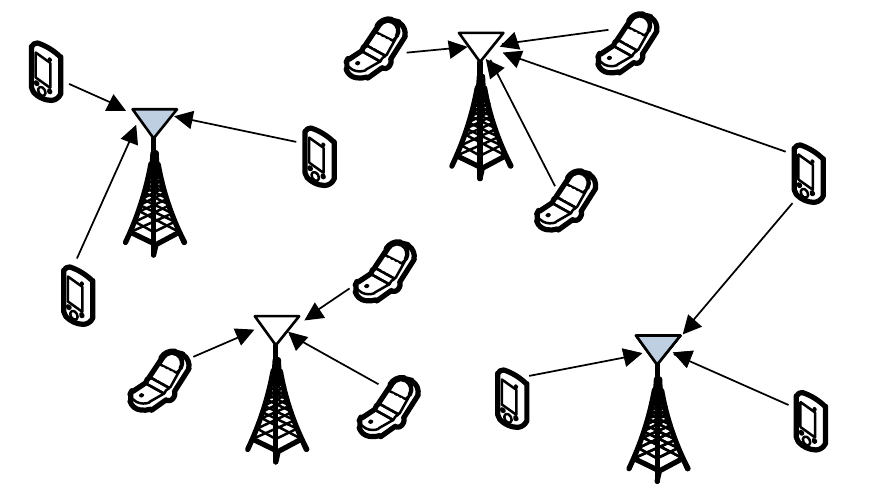}
			{\small
			\put(28,1){PBS}
			\put(15,25){SBS}
			\put(15,7){PU}
			\put(6,17){SU}
			\put(77,53){User $j$}
			\put(88,43){User $i$}
			\put(52,55){BS 1}
			\put(28,22){BS 2}
			\put(71,20){BS 3}
			\put(14,46){BS 4}
			\put(77.5,23){\rotatebox{50}{$\hup{3}{i}=\hup{b_i}{i}$}}
			\put(66,45){\rotatebox{-19}{$\hup{1}{i}=\hup{b_j}{i}$}}
			}
		\end{overpic}}
		\caption{System model for a network with two primary and two secondary BSs.} \vspace{-10pt}
		\label{fig:system_model}
	\end{figure}
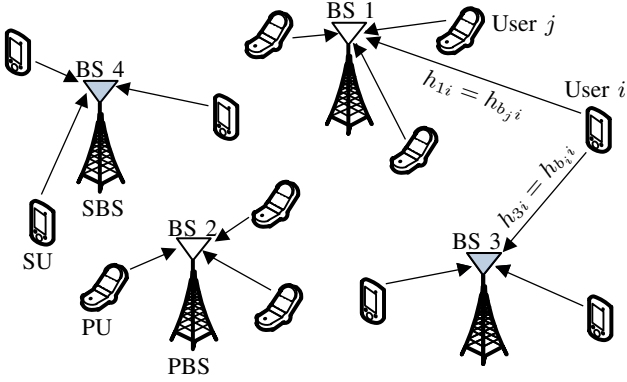

	Consider a multi-cell wireless network consisting of a set of $M=M^{p}+M^s$ users denoted by $\M=\{1,2,\cdots ,M\}$ including a set of $M^{p}$ PUs denoted by $\M^{p}=\{1,2,...,M^{p}\}$ and a set of $M^s$ SUs denoted by $\M^s=\{M^{p}+1,M^{p}+2,...,M^{p}+M^s\}$. Also, assume that there exists a set of $B=B^{p}+B^s$ base stations (BSs) denoted by $\B=\{1,2,\cdots,B\}$ including a set of $B^{p}$ primary BSs (PBSs) denoted by $\B^{p}=\{1,2,...,B^{p}\}$ serving the PUs and a set of $B^s$ cognitive radio (secondary) BSs (SBSs) denoted by $\B^s=\{B^{p}+1,B^{p}+2,...,B^{p}+B^s\}$ serving the SUs. Although the model is presented for cellular CRNs, it can also be applied to the ad-hoc CRNs by assuming that $\M^{p}$ and $\M^{s}$ denote correspondingly the set of primary and secondary links (transmitter-receiver pairs), and $B^p$ and $B^s$ denote correspondingly the set of receiving-points of the primary and secondary links \footnote{In cellular networks, at moderate or heavy traffic loads we generally have \mbox{$B \ll M$}, while in ad-hoc networks where each communication link corresponds to a unique transmitter-receiver pair, we always have $M=B$. Since the system model applies to both cellular and ad-hoc networks, the results presented  in this paper are applicable to both of these types of networks. In Section VI, numerical results  will be presented for both cellular and ad-hoc CRNs.}.
	 Fig. \ref{fig:system_model} illustrates the system model for a network consisting of two PBSs and two SBSs.
	
	Let $b_i\in\B$ denote the serving BS of user $i$ and let $\M^{p}_m$ and $\M^s_n$ denote the sub-set of PUs and SUs associated with BSs $m\in\B^{p}$ and $n\in\B^s$, respectively, i.e.,
	\begin{align*}
		\M^{p}_m &= \{i\in\M^{p} |\ \ b_i=m \}, 
	\end{align*}
	and
	\begin{align*}
		\M^s_n &= \{i\in\M^s |\ \ b_i=n \}.
	\end{align*}
	
	Let $\pup_i$ be the transmit power of user $i$ and assume that $\hup{m}{i}$, $\hup{b_i}{i}$ and $\hup{b_j}{i}$ denote the uplink path-gain from user $i$ toward BS $m\in\B$, and the corresponding BSs of users $i$ and  $j$, respectively. Let $\Nup{b_i}$ denote the noise power at the corresponding BS of user $i$, which is assumed to be additive white Gaussian. The transmit power $p_i$ is always limited to a maximum threshold denoted by $\pmax_i$ (i.e., \mbox{$p_i\in[0, \pmax_i]$}). Considering the receiver to be a conventional matched filter, for any given uplink transmit power vector $\mathbf{p}$ ($\pbold =\![p_1,p_2,...,p_M]^\mathrm{T}$), the uplink SINR of user $i$ at its BS, which is denoted by $\gamma_i$, is
	
	\begin{align}
		\label{eq:p_to_SINR}
		\gamma_i(\pbold)
		&= \dfrac	{\hup{b_i}{i} p_i}
								{ \sum\limits_{\substack{j\in\M\\ j \ne i}}{\!\!\hup{b_i}{j} p_j} \! + \!
						  \Nup{b_i}
								} \nonumber \\
		&=
		\begin{cases}
			\dfrac	{\hup{b_i}{i} p_i}
					{ \sum\limits_{\substack{j\in\M^{p}\\ j \ne i}}{\!\!\!\!\hup{b_i}{j} p_j} \! + \!
			  \sum\limits_{\substack{j\in\M^s}}{\!\!\hup{b_i}{j} p_j} +
			  \Nup{b_i}
					}, \ \mathrm{if\ } i\in\M^{p} \!\! \!\!
			\\
			\dfrac	{\hup{b_i}{i} p_i}
						{ \sum\limits_{\substack{j\in\M^{p}}}{\!\!\hup{b_i}{j} p_j} \! + \!
				  \sum\limits_{\substack{j\in\M^s\\ j \ne i}}{\!\!\!\!\hup{b_i}{j} p_j} +
				  \Nup{b_i}
						}, \ \mathrm{if\ } i\in\M^s.
		\end{cases}
	\end{align}

	Given a power vector $\pbold$, the corresponding SINR vector of the users $\gammabold=[\gamma_1,\gamma_2,...,\gamma_M]^{\mathrm{T}}$ is obtained from \eqref{eq:p_to_SINR}. We use the notations $\gammabold^p$, $\gammabold^s$, $\gammabold^p_m$, and $\gammabold^s_n$ for the SINR vector of the PUs, SINR vector of the SUs, SINR vector of the subset of PUs being served at BS $m\in\B^p$, and SINR vector of the subset of SUs being served at BS $n\in\B^s$, respectively.
	
		Given an SINR vector $\gammabold=[\gamma_1,\gamma_2,...,\gamma_M]^{\mathrm{T}}$, we can rewrite \eqref{eq:p_to_SINR} in matrix form and obtain  the corresponding power vector $\pbold$ as \cite{constrained_TPC}
		\begin{equation}
		\label{eq:gamma_to_p}
			\pbold(\gammabold)=\big(\mathbf{I}-\mathbf{F}(\gammabold)\big)^{-1}\mathbf{U}(\gammabold),
		\end{equation}
		where  $\mathbf{I}$ is an $M\times M$ identity matrix, and
		\begin{equation}
	 	\label{eq:41}
			\mathbf{U}(\gammabold)=\left[\frac{\gamma_{1} \Nup{b_1}}{\hup{b_1}{1}}, \frac{\gamma_{2} \Nup{b_2}}{\hup{b_2}{2}}, \cdots, \frac{\gamma_{M} \Nup{b_M}}{\hup{b_M}{M}} \right]^{\mathrm{T}},
		\end{equation}
		and $\mathbf{F}(\gammabold)$ is a $M\times M$ matrix with
		\begin{equation}
		\label{eq:6}
			F_{ij}=
			\begin{cases}
				0,& \mathrm{if}\quad i=j \\
				\dfrac{\gamma_i \hup{b_i}{j}}{\hup{b_i}{i}}, & \mathrm{if}\quad i\neq j.
			\end{cases}
		\end{equation}
	
	\begin{definition}
		A given uplink SINR vector $\gammabold$ is feasible if there exists a feasible transmit power vector $\pbold$ (i.e., $0\leq p_i\leq\pimax$ for all $i\in\M$) which results in $\gammabold$. In other words, $\gammabold$ is feasible if it belongs to the set of feasible SINR vectors $\mathbf{F}_{\gammabold}$, where
		\begin{align}
		\label{eq:feasible_uplink_sinr}
			\Fbold_{\gammabold}=\{\gammabold | 0\leq p_i(\gammabold) \leq \pmax_i, \ \ \forall i\in\M \}.
		\end{align}
		
	\end{definition}

\subsection{Protection Constraints for Primary Users}
	\label{sec:problem_statement}
In what follows, we define the concept of  protection for a given user and review two well-known optimization problems in underlay CRNs with  protection constraints for the PUs. Then, we discuss the need to characterize PUs' protection in such optimization problems.
	Let $\gammahat_i$ be the minimum acceptable SINR of user $i\in\M$ known as the target-SINR of the corresponding user. 
	
	\begin{definition}
			For a given power vector $\mathbf{0}\leq \pbold\leq \pbold^{\mathrm{max}}$, a user $i\in\M$ is said to be protected if $\gamma_i(\pbold)\geq \gammaihat$, where $\gamma_i(\pbold)$ is obtained from \eqref{eq:p_to_SINR}. Correspondingly, for a given SINR vector $\gammabold\geq\gammahatbold$, a user $i\in\M$ is said to be protected if $0\leq p_i(\gammabold)\leq p_i^{\mathrm{max}}$ where $p_i(\gammabold)$ is obtained from \eqref{eq:gamma_to_p}.
		\end{definition}

	
	Assuming that all PUs can be protected in the absence of the SUs, it is desirable to design a scheme to admit all or a subset of the SUs into the set of active users so that a given objective function $f_o(\pbold)$ (e.g., maximum number of protected SUs, or maximum aggregate throughput of the SUs) is optimized subject to the PUs' protection constraints and at the same time the transmit power levels for the SUs are feasible and the QoS requirements of the admitted SUs are  met. This corresponds to the following general optimization problem:
		\begin{align}
		\label{eq:opt_main}
			\underset{\pbold}{\mathrm{maximize}}  &\quad f_o(\pbold) \nonumber 
	 	\\
	 		\mathrm{subject\ to} &\quad \mathrm{protection\ of\ the\ PUs}, \nonumber  
		\\
	 		 &\quad \mathrm{feasibility\ of\ trasnmit\ power\ levels,\ and}\nonumber 
	 		 	\\
	 		  		 &\quad \mathrm{ QoS\ requirements\ for\ the\ admitted\ SUs}.
		\end{align}
		Two well-known examples of the above general optimization problem in a CRN are given bellow.
		
		\textit{Maximizing the number of protected SUs in an infeasible system:}
		In an infeasible system, there exists no feasible power vector for protecting all PUs and SUs simultaneously. Given a power vector $\pbold$, let $\s^s(\pbold)$ denote the subset of SUs that are protected, i.e., $\s^s(\pbold)=\{i\in\M^s  | \gamma_i(\pbold) \geq \gammaihat\}$.
		A feasible power vector $\pbold$ is to be obtained for which all PUs together with maximum possible number of SUs are protected. This corresponds to the solution of the following optimization problem:
		\begin{align}
			\label{eq:opt_max_supported}
				\underset{\pbold}{\mathrm{maximize}}& \quad |\s^s(\pbold)|   
		 \nonumber  \\
		 \mathrm{subject\ to} &\quad 0\leq p_i \leq \pimax, & \forall i\in\M^p, 
	 	 \nonumber \\
		 & \quad \gammai(\pbold)\geq \gammaihat,  & \forall i\in\M^p,
		 \nonumber \\
	 	& \quad 0\leq p_i \leq \pimax, & \forall i\in\M^s,
		\end{align}
		where the first two constraints correspond to the PUs' protection constraints and the last constraint corresponds to the feasibility of the SUs' transmit power levels.
	
		\textit{Maximizing the aggregate throughput of SUs in a feasible system: }
		In a feasible system, where all PUs and SUs can simultaneously be protected, it is generally desirable to provide SUs with as high SINR values as possible to maximize the aggregate throughput (aggregate channel capacity) of the SUs expressed as follows:
		\begin{align}
			\label{eq:opt_max_throughput}
			\underset{\pbold}{\mathrm{maximize}}& \ \ \ \sum_{i\in\M^s}{\log(1+\gammai(\pbold)) }    
		\nonumber \\
		\mathrm{subject\ to}\quad & 0\leq p_i \leq \pimax, & \forall i\in\M^p,
		\nonumber \\
		&  \gammai(\pbold)\geq \gammaihat,  & \forall i\in\M^p,
		\nonumber \\
		& 0\leq p_i \leq \pimax, & \forall i\in\M^s,
		\nonumber \\
		&  \gammai(\pbold)\geq \gammaihat,  & \forall i\in\M^s,
		\end{align}
		where the first two constraints correspond to the PUs' protection constraints and the last two constraints correspond to the protection of SUs.

	In general, as seen in \eqref{eq:opt_max_supported} and \eqref{eq:opt_max_throughput}, the objective function is  a function of the SINRs of SUs. It is desirable to state and solve the optimization problem at the CRN level with minimal involvement of the PRN. That is, the corresponding objective and constraint functions for the CRN should depend on the transmit power levels of the SUs and the PUs' protection constraints. These protection constraints should preferably not depend on the PRN  variables such as the  transmit power levels of the PUs. 
	But it is observed that the PUs' protection constraints at the current form expressed in the  first two constraints in \eqref{eq:opt_max_supported} and \eqref{eq:opt_max_throughput} depend on the instantaneous power vector (or correspondingly SINR vector) of PUs.
	Therefore, to avoid this coupling and decouple the optimization problem for CRN from that of PRN,
	 we	need to obtain an equivalent constraint for the PUs' protection constraints, based on the concept of FCIR. This constraint will depend on the transmit power levels of SUs and a very small amount of information pertaining to the PRN, and it will be independent of the optimization variables (power levels) related to PUs. We will derive this constraint later in this paper. This will help the CRN to guarantee the protection of PUs with minimal information feedback from the PRN. 
	
	As has been mentioned before, in the state-of-the-art interference management schemes for underlay CRNs 
			(e.g.,  \cite{JPAC_adhoc_convex_relaxation_single_PU, distributed_JPAC_antenna_arrays, 7, JPAC_SSA1,ISMIRA, LGRA,monemi_ESRPA,throughput_in_mesh_networks,bad_ITL6_IET_throughput,GP_bad_ITL,badITL4,throughput_coupled_interference_game_bad_ITL,Robust_Max_Throughput_Wang_2015,throughput_distributed_game_bad_ITL,Beamforming_Cognitive_2015,Max_Throughput_Femto_Zhang_2015 }), it is assumed that the  protection constraints for the PUs are satisfied if the cognitive interference for each primary receiving-point is lower than  the interference temperature limit (ITL) for the corresponding receiving-point.  This corresponds to a box-like FCIR, which is not really the case, as we will show in the next section.		 
	In the following two sections, we first characterize the interference feasible region, and show that it is a polyhedron (i.e.,  the maximum feasible cognitive interference threshold for each primary receiving-point is not a constant, and it depends on that for each of the other primary receiving-points). We then use it to propose two efficient power control algorithms to solve \eqref{eq:opt_max_supported} and \eqref{eq:opt_max_throughput}, respectively, which require minimal message exchange between the CRN and  the PRN.
	
\section{Characterization of Feasible Interference Region}
	\label{sec:charecterizing_FIR}
	In this section, we state the protection constraints for the PUs (the first two constraints of the optimization problems of \eqref{eq:opt_max_supported} and \eqref{eq:opt_max_throughput}) in terms of the constraints on the maximum interference that can be caused to the PBSs (which are the PU receivers for uplink cellular transmissions). For each PBS, this can be defined in two ways, namely, the \emph{total interference} caused by all SUs together with all PUs not being served by the corresponding PBS, and the \emph{cognitive interference} caused by all SUs to the corresponding PBS. Corresponding to the total and cognitive interferences, we formally introduce two different presentations for the \emph{feasible interference region} (FIR) for the PRN, namely, the \emph{feasible total interference region} (FTIR) and the \emph{feasible cognitive interference region} (FCIR). Then we derive the exact regions of FTIR and FCIR. We show that the PUs are protected if the total interference or the cognitive interference caused to the PBSs lies within the FTIR or FCIR, respectively. We  clarify the difference between these two presentations  and also describe which presentation is more preferred in designing practical power control algorithms for CRNs. 

\subsection{Protection Constraints for PUs Based on FTIR}
	Given the uplink power vector $\pbold$ corresponding to an SINR vector $\gammabold$, let $I^p_m$ denote the total interference caused by all SUs together with PUs $i\notin \M^p_m$ on the PBS $m\in\B^p$, i.e.,
	\begin{align}
		\label{eq:231}
		I^p_m 
		&=
			\sum_{\substack{k\in\B \\ k\neq m}}\sum_{i\in\M_k} \!\! {(p_i h_{mi})} \nonumber \\
		&= \sum_{\substack{k\in\B^p \\ k\neq m}}\sum_{i\in\M_k} \!\!{(p_i h_{mi})} + \!\! \sum_{k\in\B^s}\sum_{i\in\M_k}\!\!{(p_i h_{mi})},
		\ \forall m\in\B^p.
	\end{align}
	\begin{lemma}
		For each PBS $m\in\B^p$, given the SINRs of the PUs in the corresponding cell (i.e., $\gammabold^p_m$) and the total interference caused by other cells to that cell (i.e., $I^p_m$), the transmit power of each user $i\in\M_m$, denoted by $p_i(\gammabold^p_m,I^p_m)$, is obtained as
		\begin{align}
			\label{eq:pi_FTIR}
			p_i(\gammabold^p_m,I^p_m)= \frac{1}{h_{mi}}\!\times\!\dfrac{\gamma_i}{(\gamma_i+1)} \left( \dfrac{ I^p_m + N_m} {1- \! \sum\limits_{j\in\M_m}\!\!\! \frac{\gamma_j}{\gamma_j+1}} \right).
		\end{align}
	\end{lemma}
	\begin{IEEEproof}
		Let $\Phi_{n}$ denote the total received power plus noise at the BS $n\in\B$, i.e., $\Phi_{n}=\sum_{j\in\M}{\!\hup{n}{j} p_j} \!  + \Nup{n}$. Thus $\Phi_{b_i}$ represents the total received power plus noise at the BS serving user $i$, i.e., 
		\begin{align}
		\label{eq:phi_i229}
			\Phi_{b_i}=\sum\limits_{j\in\M}{\!\!\hup{b_i}{j} p_j} \!  + \Nup{b_i}.
		\end{align}
		From {\eqref{eq:p_to_SINR}} and {\eqref{eq:phi_i229}}, we have 
		\begin{equation}
		\label{eq:1129}
			\gammai=
			\dfrac	{\hup{b_i}{i} p_i}
					{\Phi_{b_i}-\hup{b_i}{i} p_i}, \ \forall i\in\M.
		\end{equation}
		This results in
		\begin{align}
		\label{eq:pi_versus_phi}
			p_i= \dfrac{\gammai}{(\gammai+1)} \dfrac{\Phi_{b_i}}{\hup{b_i}{i}}. 
		\end{align}
		From {\eqref{eq:pi_versus_phi}}, for each $m\in\B$ and $n\in\B$, the following is obtained:
		\begin{align}
		\label{eq:2229}
			\sum\limits_{i\in\M_{n}}{p_i \hup{m}{i}} = \Phi_{n} \sum\limits_{i\in\M_{n}}{\dfrac{\hup{m}{i}}{\hup{n}{i}}\frac{\gamma_i}{\gamma_i+1}}.
		\end{align}
		By letting $m=n$ and adding $\sum_{i\notin\M_{m}}{\!p_i \hup{m}{i}}+N_{m}$ to both sides of {\eqref{eq:2229}}, $\Phi_{m}$ is obtained as
		\begin{align}
		\label{eq:2349}
			\Phi_{m}
			= & \dfrac{\sum\limits_{i\notin\M_m} {\!\!\!\big(p_i \hup{m}{i} \big)} + N_{m}} {1- \! \sum\limits_{i\in\M_m}\!\!\! \left(\frac{\gamma_i}{\gamma_i+1}\right)}, \ \forall m\in\B.
		\end{align}
		From \eqref{eq:pi_versus_phi} and \eqref{eq:2349} and the fact that $I^p_m=\sum_{i\notin\M_m} {\!(p_i \hup{m}{i} )}$, \eqref{eq:pi_FTIR} is concluded.
	\end{IEEEproof}
	\begin{definition}
	\label{def:total_feasible_interference}
		Let $\mathbf{I}^{p}=[I^{p}_1,I^{p}_2,\cdots,I^{p}_{B^p}]^{\mathrm{T}}$ denote the vector of total interference  caused to the PBSs. Given the SINR vector of PUs $\gammabold^{p}\geq \gammahatbold^{p}$, the feasible total interference region (FTIR) $\mathbf{F}^{p}_{\mathbf{\mathcal{I}}}\subset \mathbb{R}_{+}^{B^p}$ is defined as the vector space of total interference caused to the PBSs for which all the PUs are protected, i.e., 
		\begin{align}
		\label{eq:320}
			\mathbf{F}^{p}_{\mathbf{\mathcal{I}}}=\{\mathbf{I}^{p} | 0 \leq p_i\left(\! \gammabold^p,{I}^{p}_{m} \! \right) \! \leq \! p_i^{\mathrm{max}}, \ \forall i\in\M_m, \forall m\in\B^p \},
		\end{align}
		where $p_i\left( \gammabold^p,{I}^{p}_{m} \right)$ is obtained from \eqref{eq:pi_FTIR}. Besides, given $\gammabold^p$, we say that $\mathbf{I}^{p}$ is feasible if $\mathbf{I}^{p}\in \mathbf{F}_{\mathcal{I}}^{p}$.
	\end{definition}
	The following corollary is directly obtained from \eqref{eq:pi_FTIR}.
	
	\newcommand{\minp}{\min\limits_{i\in\M_m}\!\!\left(\frac{p_i^{\mathrm{max}} h_{mi}(\gamma_i+1)}{\gamma_i}\right)}
	\begin{corollary}
		\label{col:FTIR}
		The FTIR is a closed box of the following form:
		\begin{align}
		\label{th:FTIR}
			0\leq I^{p}_m \leq \overline{I}^p_m, \ \ \forall m\in\B^p,
		\end{align}
		where $\overline{I}^p_m$ is called the total interference temperature limit (TITL) of the PBS $m\in\B^p$ and is obtained as
		\begin{align}
		\label{eq:ITL3}
			\overline{I}^p_m \! = \! \minp 
			\! \times \!
			\bigg( 1- \!\!\! \sum_{j\in\M_m}\!\frac{\gamma_j}{\gamma_j+1} \bigg) - N_m.
		\end{align}
	\end{corollary}

\subsection{Protection Constraints for PUs Based on FCIR}

	Given the uplink power vector $\pbold$ corresponding to an SINR vector $\gammabold$, let $I^{s\rightarrow p}_m$ denote the interference caused by all of the SUs to the PBS $m\in\B^p$, referred to as the cognitive interference caused to the PBS $m\in\B^p$. That is, 
	\begin{align}
	\label{eq:I_s_to_p}
		I^{s\rightarrow p}_m=\sum_{i\in\M^s}{(p_i h_{mi})},\ \ \ \forall m\in\B^p.
	\end{align}


	
	In what follows, given the SINRs of the PUs and the vector corresponding to cognitive interference (referred to as cognitive interference vector) imposed on the primary receiving-points, we derive the corresponding power vector of the PUs, based on which, we are able to obtain the constraint leading to the FCIR.
	\begin{theorem} 
		Given the SINR vector of the PUs $\gammabold^p$ and the cognitive interference vector caused to the PBSs by the SUs ($\mathbf{I}^{s\rightarrow p}=[I^{s \rightarrow p}_1,I^{s \rightarrow p}_2,\cdots,I^{s \rightarrow p}_{B^p}]^{\mathrm{T}}$), the transmit power for each PU $i\in\M^p$ (corresponding to  $\gammabold^p$ and $\mathbf{I}^{s\rightarrow p}$), denoted by $p_i(\gammabold^p,\mathbf{I}^{s\rightarrow p})$, is obtained as 
		\begin{align}
		\label{eq:p_s_2_p}
			p_i(\gammabold^p,\mathbf{I}^{s\rightarrow p})= \dfrac{\gamma_i}{(\gamma_i+1)} 
				\! \times \! 
			\dfrac{ \Phi_{b_i} \! \left(\gammabold^p,\mathbf{I}^{s\rightarrow p}\right)}{\hup{b_i}{i} } ,  \ \forall i\in\M^p,
		\end{align}
		where $\Phibold^p=[\Phi_{1}, \Phi_{2}, ..., \Phi_{B^p}]^{\mathrm{T}}$ is obtained as
		\begin{align}
		\label{eq:phi_s_2_p}
	 		\Phibold^p\left( \gammabold^p,\mathbf{I}^{s \rightarrow p} \right)= \left( \mathbf{I-H}(\gammabold^p) \right)^{-1} \left( \mathbf{N}^p + \mathbf{I}^{s \rightarrow p} \right),
		\end{align}
		in which $\mathbf{I}$ is a $B^p\times B^p$ identity matrix, $\mathbf{N}^p=[N_{1}, N_{2}, ..., N_{B^p}]^{\mathrm{T}}$, and $\mathbf{H}$ is a $B^p\times B^p$ matrix whose elements are given by
		\begin{align}
		\label{eq:H_elements2}
			H_{mn}=
			\begin{cases}
			\sum\limits_{i\in\M_{m}}{\!\! \frac{\gamma_i}{\gamma_i+1}}, & \mathrm{if\ } m=n, \\
			\sum\limits_{i\in\M_{n}}{\!\! \frac{\hup{m}{i}}{\hup{n}{i}} \frac{\gamma_i}{\gamma_i+1} }, & \mathrm{if\ } m\neq n.
			\end{cases}
		\end{align}
	\end{theorem}
	\begin{IEEEproof}
		From \eqref{eq:2349} and \eqref{eq:2229}, for any PBS $m\in\B^p$, we have
		\begin{align}
		\label{eq:234}
			\Phi_{m}
					= & \dfrac{\sum\limits_{i\notin\M_m} {\!\!\!\big(p_i \hup{m}{i} \big)} + N_{m}} {1- \! \sum\limits_{i\in\M_m}\!\!\! \left(\frac{\gamma_i}{\gamma_i+1}\right)}
					\nonumber \\
			= &  \dfrac{ \sum\limits_{\substack{n\in\B^p\\ n\neq m}} \!\!\! \sum\limits_{\ \ i\in\M_n} {\!\!\!\big(p_i \hup{m}{i} \big)} + \sum\limits_{\substack{n\in\B^s}} \!\!\! \sum\limits_{\ \ i\in\M_n} {\!\!\!\big(p_i \hup{m}{i} \big)} + N_{m}}
			  {1- \! \sum\limits_{i\in\M_m}\!\!\! \left( \frac{\gamma_i}{\gamma_i+1} \right)}
			\nonumber \\
			 = & \dfrac{\sum\limits_{\substack{n\in\B^p\\ n\neq m}} \!\!\! \Phi_{n} \sum\limits_{i\in\M_{n}}{ \frac{\hup{m}{i}}{\hup{n}{i}} \frac{\gamma_i}{\gamma_i + 1} } +I_m^{s \rightarrow p} + N_{m}} {1- \! \sum\limits_{i\in\M_m}\!\!\! \left( \frac{\gamma_i}{\gamma_i+1} \right)}, \ \forall m\in\B^p.
		\end{align}
		From \eqref{eq:234}, we conclude that
		\begin{align}
			\label{eq:255}
			\Phi_{m} \left( 1- \! \sum\limits_{i\in\M_m}\!\!\! \left( \frac{\gamma_i}{\gamma_i+1} \right) \right) -
			\sum\limits_{\substack{n\in\B^p\\ n\neq m}} \ \! \Phi_{n} \! \sum\limits_{i\in\M_{n}}{ \! \dfrac{\hup{m}{i}}{\hup{n}{i}}\ \!\left(\frac{\gamma_i}{\gamma_i+1}\right) } \nonumber \\
			= I^{s \rightarrow p}_m + N_{m},  m=1,2,...,B^p.
		\end{align}
		Writing \eqref{eq:255} in matrix form results in \eqref{eq:phi_s_2_p} which together with \eqref{eq:pi_versus_phi} results in \eqref{eq:p_s_2_p}.
	\end{IEEEproof}
	\begin{definition}
	\label{def:feasible_interference}
		Given the SINR vector of PUs $\gammabold^{p}\geq \gammahatbold^p$, the feasible cognitive interference region (FCIR) $\mathbf{F}^{s\rightarrow p}_{\mathbf{\mathcal{I}}}\subset \mathbb{R}_{+}^{B^p}$ is defined as the space of interference vectors caused by the CRN (i.e., SUs) to the PBSs for which all the PUs are protected, i.e., 
		\begin{align}
		\label{eq:32}
			\mathbf{F}^{s\rightarrow p}_{\mathbf{\mathcal{I}}} \! = \! \{\mathbf{I}^{s\rightarrow p} | 0 \leq p_i\left(\! \gammabold^p,\mathbf{I}^{s\rightarrow p} \! \right) \! \leq \! p_i^{\mathrm{max}} \! , \forall i\in\M_m, \forall m\in\B^p \},
		\end{align}
		where $p_i\left( \gammabold^p,\mathbf{I}^{s\rightarrow p} \right)$ is obtained from \eqref{eq:p_s_2_p}. Furthermore, we say that a given cognitive interference vector $\mathbf{I}^{s\rightarrow p}$ is feasible if $\mathbf{I}^{s\rightarrow p}\in \mathbf{F}_{\mathcal{I}}^{s\rightarrow p}$.
	\end{definition}
	
	In the following theorem, we derive the FCIR.
	\begin{theorem}
	\label{th:FCIR}
		The FCIR is a polyhedron given by the following matrix inequalities:
	\begin{align}
	\label{eq:321}
		\mathbf{0} \leq \mathbf{I}^{s\rightarrow p},
	\end{align}
	and
	\begin{align}
	\label{eq:322}
		\mathbf{A}^{\!p}  \mathbf{I}^{s\rightarrow p}  
		\leq \mathbf{C}^p,
	\end{align}
		where 
		\begin{align}
		\label{eq:A_feas_coef}
			\mathbf{A}^{p}&=\left( \mathbf{I-H}(\gammabold^p) \right)^{-1},
		\end{align}
		and
		\begin{align}
		\label{eq:B_feas_coef}
			\mathbf{C}^{p}&=\Phibold^{p,\mathrm{max}} - \left( \mathbf{I-H}(\gammabold^p) \right)^{-1}  \mathbf{N}^p,
		\end{align}
		in which $\mathbf{I}$ is a $B^p \times B^{p}$ identity matrix and $\Phibold^{p,\mathrm{max}}=[\Phi_1^{\mathrm{max}},\Phi_2^{\mathrm{max}},\cdots,\Phi_{B^{p}}^{\mathrm{max}}]^{\mathrm{T}}$,
		where $\Phi^{\mathrm{max}}_m= \min\limits_{i\in\M_{m}} {\{p_i^\mathrm{max} \hup{m}{i}} \frac{\gamma_i+1}{\gamma_i}\}$.
	\end{theorem}
	\begin{IEEEproof}
		Since we have assumed that the feasibility of the system holds in the absence of SUs (i.e., when $I^{s\rightarrow p}_m=0, \forall m\in\B^{p}$), we have $\left. 0\leq p_i(\gammabold^{p},\mathbf{I}^{s \rightarrow p})\right|_{ \mathbf{I}^{s\rightarrow p} = \mathbf{0}}$. Therefore, from \eqref{eq:p_s_2_p} and \eqref{eq:phi_s_2_p}, we conclude that $\left( \mathbf{I-H}(\gammabold^{p}) \right)^{-1}$ exists and from Perron-Frobenius theorem \cite{perron}, it is positive component-wise. The feasibility of $\mathbf{I}^{s\rightarrow p}$ leads to the protection of all PUs for all $m\in\B^{p}$. Therefore, from Definition \ref{def:feasible_interference} and from \eqref{eq:p_s_2_p} and \eqref{eq:phi_s_2_p}, we conclude that
		\begin{align}
			\label{eq:55}
			\mathbf{0}\leq \left( \mathbf{I-H}(\gammabold^{p}) \right)^{-1} \left( \mathbf{N}^{p} + \mathbf{I}^{s \rightarrow p} \right) \leq \Phibold^{p,\mathrm{max}},
		\end{align}
		where $\Phibold^{p,\mathrm{max}}$ is given in the declaration of the proposition. The right inequality of \eqref{eq:55} directly results in \eqref{eq:322}. The left hand inequality of \eqref{eq:55} together with the fact that the feasible cognitive interference imposed on any PBS is non-negative results in $\max\{-\left( \mathbf{I-H}(\gammabold^{p}) \right)^{-1}  \mathbf{N}^{p},\mathbf{0}\}\leq  \mathbf{I}^{s \rightarrow p} $ which leads to \eqref{eq:321}. 
		This completes the proof.
	\end{IEEEproof}
	
	Theorem \ref{th:FCIR} derives an equivalent constraint for PUs' protection in terms of FCIR given by \eqref{eq:321} and \eqref{eq:322}.
	In other words, the PUs' protection constraints expressed as  $0\leq p_i \leq \pimax$ and $\gammai(\pbold)\geq \gammaihat$, for all $i\in\M^p$ (for example in the optimization problems \eqref{eq:opt_max_supported} and \eqref{eq:opt_max_throughput}) can be now replaced with the constraints \eqref{eq:321} and \eqref{eq:322}. Note that in the FCIR characterized by \eqref{eq:321} and \eqref{eq:322}, the cognitive interference vector $\mathbf{I}^{s\rightarrow p}$ is a function of the transmit power level for SUs (as defined in \eqref{eq:I_s_to_p}) and the coefficients of $\mathbf{A}^p$ and $\mathbf{C}^p$ are functions of $\gammabold^p$ (as seen in \eqref{eq:A_feas_coef} and \eqref{eq:B_feas_coef}). Given $\gammabold^p$, the explicit instantaneous transmit power levels of the users are not required for the calculation of these parameters. This enables us to state the interference optimization problems at the CRN level with minimal involvement of the PRN, i.e., only the coefficients $\mathbf{A}^p$ and $\mathbf{C}^p$, which are functions of the parameter values pertinent to the PRN (as seen in \eqref{eq:A_feas_coef} and \eqref{eq:B_feas_coef}), are required to be given by the PRN to the CRN. 
	
\subsection{Example and Discussion}
\label{sec:charecterizing_FIR_eg_disc}
	In the following, we first illustrate an instance of a two-dimensional FTIR and FCIR for a simple PRN consisting of two PBSs coexisting with a CRN and then we discuss which of the FTIR or FCIR is better to be used to express the PUs' protection requirements in practical resource allocation problems for CRNs. 
	
	\emph{Example 1:} Consider a CRN wherein the primary tier consists of two BSs ($\B^{p}=\{1,2\}$). From Corollary \ref{col:FTIR}, the FTIR is the box enclosed by the following inequalities:
	\begin{align}
		0\leq I^p_1 \leq \overline{I}^p_1, \quad  \nonumber \\
		0\leq I^p_2 \leq \overline{I}^p_2, \quad  \nonumber 
	\end{align}
	where $I^p_1$ and $I^p_2$ are obtained from \eqref{eq:ITL3} and from Theorem \ref{th:FCIR}, the FCIR is the space enclosed by the following inequalities:
	\begin{align}
		& A_{11}^p I^{s \rightarrow p}_1 + A_{12}^p I^{s \rightarrow p}_2 \leq C_1^p, \nonumber \\
		& A_{21}^p I^{s \rightarrow p}_1 + A_{22}^p I^{s \rightarrow p}_2 \leq C_2^p, \nonumber \\
		& 0 \leq I^{s \rightarrow p}_1,	\nonumber \\
		& 0 \leq I^{s \rightarrow p}_2,	\nonumber
	\end{align}
	where the coefficients $\mathbf{A}^p$ and $\mathbf{C}^p$ are obtained from \eqref{eq:A_feas_coef} and \eqref{eq:B_feas_coef}, respectively, as follows:
	
	\begin{align*}
		\begin{pmatrix}
		  A_{11}^{p} & A_{12}^{p} \\
		  A_{21}^{p} & A_{22}^{p} 
		 \end{pmatrix}
		 = \frac{1}{K(\gammabold^p)} \times  
		 \begin{pmatrix}
			  1-\!\!\sum\limits_{i\in\M_{2}}{\!\!\!\! \frac{\gamma_i}{\gamma_i+1}} & 
			  {\sum\limits_{i\in\M_{2}}{\!\! \frac{h_{1i}}{h_{2i}} \frac{\gamma_i}{\gamma_i+1}}} \\
			  {\sum\limits_{i\in\M_{1}}{\!\! \frac{h_{2i}}{h_{1i}} \frac{\gamma_i}{\gamma_i+1}}} & 
			  {1-\!\!\sum\limits_{i\in\M_{1}}{\!\! \frac{\gamma_i}{\gamma_i+1}}} 
 		 \end{pmatrix},
	\end{align*}
	and
	\begin{align*}
		\begin{pmatrix}
		 	C_1^{p} \\
	  		C_2^{p} 
		\end{pmatrix}
		=
		\begin{pmatrix}
			 \Phi_1^{\mathrm{max}} - A_{11}^{p} N_1 - A_{12}^{p} N_2 \\
			  \Phi_2^{\mathrm{max}}- A_{21}^{p} N_1-A_{22}^{p} N_2 
	 	 \end{pmatrix},
		\end{align*}
		where
		\begin{align*}
			K(\gammabold^p)= \left({1-\sum\limits_{i\in\M_{2}}{\!\! \frac{\gamma_i}{\gamma_i+1}}} \right)
							 \left({1-\sum\limits_{i\in\M_{1}}{\!\! \frac{\gamma_i}{\gamma_i+1}}} \right)
							 - \\
							 \left({\sum\limits_{i\in\M_{1}}{\!\! \frac{h_{2i}}{h_{1i}} \frac{\gamma_i}{\gamma_i+1}}} \right)
					 		 \left({\sum\limits_{i\in\M_{2}}{\!\! \frac{h_{1i}}{h_{2i}} \frac{\gamma_i}{\gamma_i+1}}} \right).
		\end{align*}
	\begin{figure}
		\centering
		\includegraphics [width=250pt]{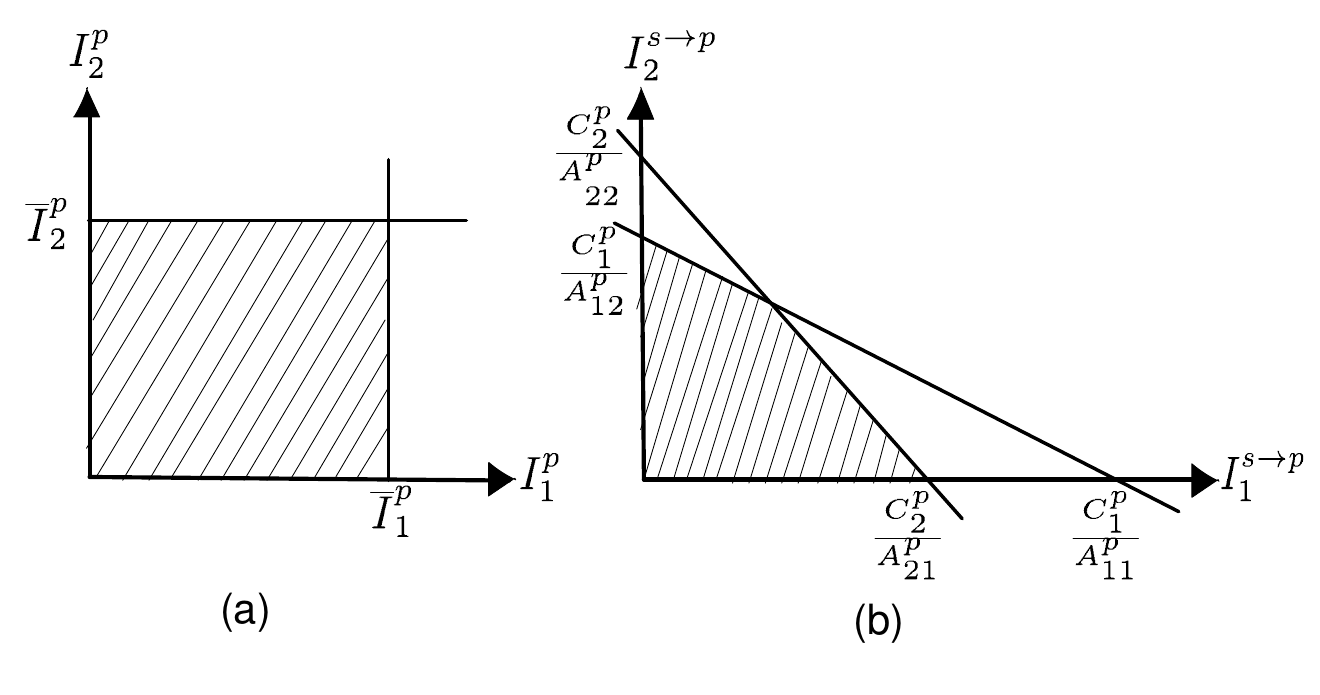}
		\vspace{-15pt}
		\caption{(a) FTIR and (b) FCIR of a CRN having two PBSs.} \vspace{-10pt}
		\label{fig:ex_FTIR_FCIR}
	\end{figure}
	Fig. \ref{fig:ex_FTIR_FCIR} shows the FTIR and FCIR of this example.
	
	
	\begin{remark}
	\label{rmk:bad_concept}
		As has been mentioned before, in the relevant literature, the cognitive interference imposed by the SUs on each of the primary receiving-points is considered to be smaller than a  value (i.e., ITL), which is assumed to be constant and independent of ITL values for other primary receiving-points.  This  is considered as the equivalent constraint for the  PUs' protection constraints. This corresponds to the assumption of a box-like FCIR (see Fig. \ref{fig:ex_TLFCIR}) which is not the case as shown in Theorem \ref{th:FCIR} and seen in the example given above.  Consequently, in the existing algorithms, the allowable cognitive interference of the SUs on primary receiving-points  is adjusted based on an incorrect assumption. 
		
		
			To elaborate, since the SUs cause interference to primary receiving points,  the SINRs at the PUs' receivers decrease when SUs are admitted into the set of active users. To ensure this does not violate the PUs' protection constraints, power control schemes must be employed at the PRN and CRN levels. At the PRN side, PUs should  employ a power control scheme and increase their transmit powers to compensate for the cognitive interference imposed by the SUs so that the desired SINRs can be achieved at the PUs' receivers. At the CRN side, it is necessary the SUs' transmit  power levels be controlled in a way that the cognitive interference lies within the characterized polyhedron region as discussed before. Therefore, the conventional assumption of fixed ITL values does not guarantee the protection of PUs unless the box formed from the fixed ITLs is placed entirely within the polyhedron corresponding to the FCIR (e.g., the box may be inscribed in the FCIR). In this way, the performance may substantially decrease (as will be seen later in the simulation results) since the inscribed box-like region does not fully utilize the whole polyhedron feasible region of the cognitive interference. Besides, since there exist innumerable boxes that may be inscribed in a polyhedron, the question on which ITL-box corresponds to better performance is an open problem. In this context, the usefulness of the polyhedron characterization of FCIR  becomes evident as well.

	\end{remark}

	\begin{remark}
		\label{rmk:1}
		As has been mentioned before, the protection of PUs can be checked by observing whether the total interference or cognitive interference caused to the PBSs lie within the FTIR or FCIR, respectively. Now, we discuss which one is  preferable from practical point of view, based on the following criteria:

		\begin{itemize}
			\item The constraint should enable us to define the (power) optimization problem (e.g., \eqref{eq:opt_max_supported} or \eqref{eq:opt_max_throughput}) at the CRN level.
			\item The constraint should impose minimal signalling overhead between the PRN and the CRN.
			
		\end{itemize}
		
		As for the former criterion, for the FTIR, the total interference vector $\mathbf{I}^p$ expressed in \eqref{eq:231}, depends on the instantaneous transmit power of the primary and secondary users. Since the value of the transmit power of the PUs may not be available to the CRN, the admission control of the SUs through comparing the total interference of the PBSs with their corresponding TITL values may only be accomplished through centralized algorithms wherein all information about the PRN and CRN are available. On the other hand, for the FCIR, the cognitive interference caused to the PBSs obtained by \eqref{eq:I_s_to_p} depends on the information available to the CRN (i.e., power level of the SUs and the path-gains between the SUs and PBSs).
		 
		As for the latter criterion, for the FCIR, the coefficients $\mathbf{A}^p$ and $\mathbf{C}^p$ in \eqref{eq:A_feas_coef} and \eqref{eq:B_feas_coef} (whose dimensions are $B^p\times B^p$ and $B^p \times 1$, respectively) for checking the feasibility of a given cognitive interference vector may be calculated by the PRN based on the information pertinent to the PUs and PBSs. Therefore, instead of providing the CRN with a large amount of information pertinent to the PRN (e.g., target-SINRs and path-gains of all PUs), the PRN may provide the CRN with information about the coefficients $\mathbf{A}^p$ and $\mathbf{C}^p$ (total of $B^p \times (B^p+1)$ real numbers) only. Subsequently, the CRN may check whether the admitted SUs guarantee the protection of PUs with low-complexity calculations through \eqref{eq:322}  without the need for detailed information of the PUs. This motivates us to devise low-complexity admission and power control algorithms as will be presented in the following section.
	\end{remark}

\section{Proposed Joint Power and Admission Control (JPAC) Algorithms Based on the Characterized FCIR}
	\label{sec:proposed_algorithm}

	We have shown that, in practical interference management algorithms,  it is not proper to consider a constant and independent ITL value for each of the primary receiving-points. This is because, using constant ITL values may result in reduced system performance and/or violation of the protection constraints for the PUs. This finding would significantly affect
	the design of practical interference management schemes for CRNs. To show this effect, based on the characterized FCIR, we consider two problems stated in \eqref{eq:opt_max_supported} and \eqref{eq:opt_max_throughput} for infeasible and feasible cases, respectively. For infeasible systems, by considering the characterized FCIR, we propose a novel JPAC algorithm to address the problem of maximizing the number of SUs. Furthermore, to show how the characterized FCIR affects the performance of power control algorithms for feasible systems, we revise one of the existing algorithms to obtain the maximum aggregate throughput of SUs. We will show that for the two distinct objectives (i.e., maximization of the number of SUs and maximization of the aggregate throughput of SUs), our proposed interference management schemes outperform the existing ones.  Also, we will show that when a box-like ITL is considered, the PUs' protection is not always guaranteed.

\subsection{Infeasible System Case: JPAC Aiming at Maximizing the Number of SUs}
	
	For an infeasible system, based on the optimization problem \eqref{eq:opt_max_supported}, we are interested to protect all PUs and support the maximum possible number of SUs. While there may exist many power vectors corresponding to the solutions of \eqref{eq:opt_max_supported}, we are interested in the ones with the minimum aggregate transmit power of the users. The following JPAC optimization problem obtains the solution corresponding to the \emph{minimal} transmit power vector \cite{monemi_ESRPA}:
	\begin{align}
			\label{eq:opt_max_supported2}
				\underset{\pbold}{\mathrm{maximize}}& \quad |\s^s(\pbold)|   
		 \nonumber  \\
		 \mathrm{subject\ to} &\quad 0\leq p_i \leq \pimax, & \forall i\in\M^p
	 	 \nonumber \\
		 & \quad \gammai(\pbold)= \gammaihat,  & \forall i\in\M^p
		 \nonumber \\
	 	& \quad 0\leq p_i \leq \pimax, & \forall i\in\M^s
	 	\nonumber \\
	 	& \quad \gammai(\pbold)\in \{ \gammaihat,0\}, & \forall i\in\M^s.
	\end{align}
		
	The optimization problem in \eqref{eq:opt_max_supported2} is a \emph{mixed-integer non-linear program} that searches through the space of SINR-vectors $\Gammabold=\prod_{i=1}^{M^p}{\{\gammahat_i\}} \times \prod_{i=M^p+1}^{M^p+M^s} \{\gammahat_i,0\}$ and finds a feasible SINR vector $\gammabold^{*}\in\Gammabold$ (and its corresponding power vector $\pbold^{*}\in [\mathbf{0}\ \mathbf{\pmax}]$) that maximizes the number of protected SUs. 
	Given the SINR vector $\gammabold\in\Gammabold$, for any SU $i\in\M^s$, $\gammai=\gammahat_i$ means that user $i\in\M^s$ is admitted into the set of protected SUs and $\gamma_i=0$ implies that the corresponding user is removed (i.e., turned off). In what follows, we propose an efficient and low-complexity algorithm to find a sub-optimal solution of \eqref{eq:opt_max_supported2}.
	
	In \cite{constrained_TPC}, the authors propose a constrained iterative target-SINR tracking power control (Constrained-TPC) algorithm in which each active user $i\in\M$ at each time-step $n+1$ updates its transmit power $p_i^{(n+1)}$ as follows:
	\begin{align}
	\label{eq:constrained_TPC}
		\textrm{Constrained-TPC:\ } p_i^{(n+1)} \!\! = \! \min \!\left\{\pmax_i,p_i^{(n)}  \frac{\gamma_i(\pbold^{(n)})}{\gammaihat} \! \right\}\!.
	\end{align}
	It is shown that the constrained-TPC always converges to a stationary power vector $\pboldzero$ wherein if the system is feasible, all users are protected by attaining their target-SINRs ($\gamma_i(\pboldzero)=\gammahat_i$, for all $i\in\M$). For the case of an infeasible system, we have $\pzero_i= p_i^\mathrm{max}$ for any non-protected user $i$. We propose our JPAC algorithm as a procedure in which all SUs are initially admitted ($\A^s\leftarrow \M^s$). Then at each step, all PUs together with active SUs update their transmit powers according to constrained-TPC  until reaching the stationary power vector. Then, one of the SUs is removed for the case where the system is infeasible. This continues until the remaining set of PUs together with active SUs are protected. Here, the key issue is to construct a	removal criterion which results in good overall performance. We	consider the following cases in each removal step.
	
	\textit{Case 1: All PUs are protected but the QoS constraints of some SUs are violated.}

	This corresponds to the case where the cognitive interference vector $\CIzerobold$ lies within the FCIR but the stationary SINRs of some SUs are smaller than the their target values. Here the primary-tier is protected and thus we need an intra-tier removal mechanism for the CRN to remove some unsupported SUs.
	
		Among all the active SUs $\A^s$, we choose the removal candidate SU $i^{*}$ as the one which has the worst ``disturbing" effect on the SBS $m^{*}\in\B^s$ with the maximum number of non-supported users, i.e.,
	\begin{align}
	\label{eq:23}
		i^{*}=\argmax_{i\in\A^s} \{ \pzero_i h_{m^{*}i} \},
	\end{align}
	where 
	\begin{align}
	\label{eq:24}
		m^{*}=\argmax_{m\in\B^s} \big| \left( \M^s \setminus \s^s(\pbold)\right)
		 \cap \M_m \big|,
	\end{align}
	in which $|.|$ is the cardinality of the corresponding vector.
	
	\textit{Case 2: The protection constraint is violated for at least one PU.}
	
	This corresponds to the case where the cognitive interference vector $\CIzerobold$ does not lie within the FCIR. Let us define $\mathbf{S}^{\mathrm{inf}}(\CIbold)$ as
	\begin{align}
	\label{eq:sm}
		\mathbf{S}^{\mathrm{inf}}(\CIbold)=
		\mathbf{A}^{\!p}  \mathbf{I}^{s\rightarrow p} - \mathbf{C}^p.
	\end{align}
	Note that, from \eqref{eq:322}, it can be seen that $\mathbf{S}^{\mathrm{inf}}(\CIbold)=\mathbf{0}$ presents the infeasibility boundaries of FCIR for the protection of PUs. In other words, $S_m^{\mathrm{inf}}(\CIbold)\leq 0$ (where $S_m^{\mathrm{inf}}$ is the $m$th element of $\mathbf{S}^{\mathrm{inf}}$) means that all PUs associated with the PBS $m\in\B^p$ are protected and $S_m^{\mathrm{inf}}(\CIbold) > 0$ means that there exists at least one unprotected PU $i\in\M_m$.
	Given the cognitive interference vector $\CIzerobold\geq \mathbf{0}$, we define the infeasibility measure of $\CIzerobold$ with respect to the PBS $m\in\B^p$ denoted by $d(\CIzerobold,S_m^{\mathrm{inf}})$ as the minimum signed Euclidean distance of $\CIzerobold$ from the cognitive interference infeasibility boundary $S_m^{\mathrm{inf}}(\CIbold)=0$. This is obtained as
	\begin{align}
	\label{eq:d}
		d(\CIzerobold,S_m^{\mathrm{inf}})= 
		\frac{\sum_{n\in\B^p} A_{mn}^p \CIzero_n - C_m^{p}}{\sqrt{\sum_{n\in\B^p} \left(A_{mn}^p\right)^2}}.
	\end{align}
	It is seen that the protection of PUs is guaranteed if  we have $S_m^{\mathrm{inf}}(\CIzerobold) \leq 0$, or equivalently, $d(\CIzerobold,S_m^{\mathrm{inf}})\leq 0$ for all $m\in\B^p$. 
	\begin{definition}
		We say that a given cognitive interference vector $\mathring{\mathbf{I}}^{s \rightarrow p}$ is feasible for BS $m\in\B^p$ if $S_m^{\mathrm{inf}}(\CIzerobold) \leq 0$, or equivalently, $d(\CIzerobold,S_m^{\mathrm{inf}})\leq 0$.
	\label{def:23}
	\end{definition}

\begin{figure*}
		\centering
		\includegraphics [width=280pt]{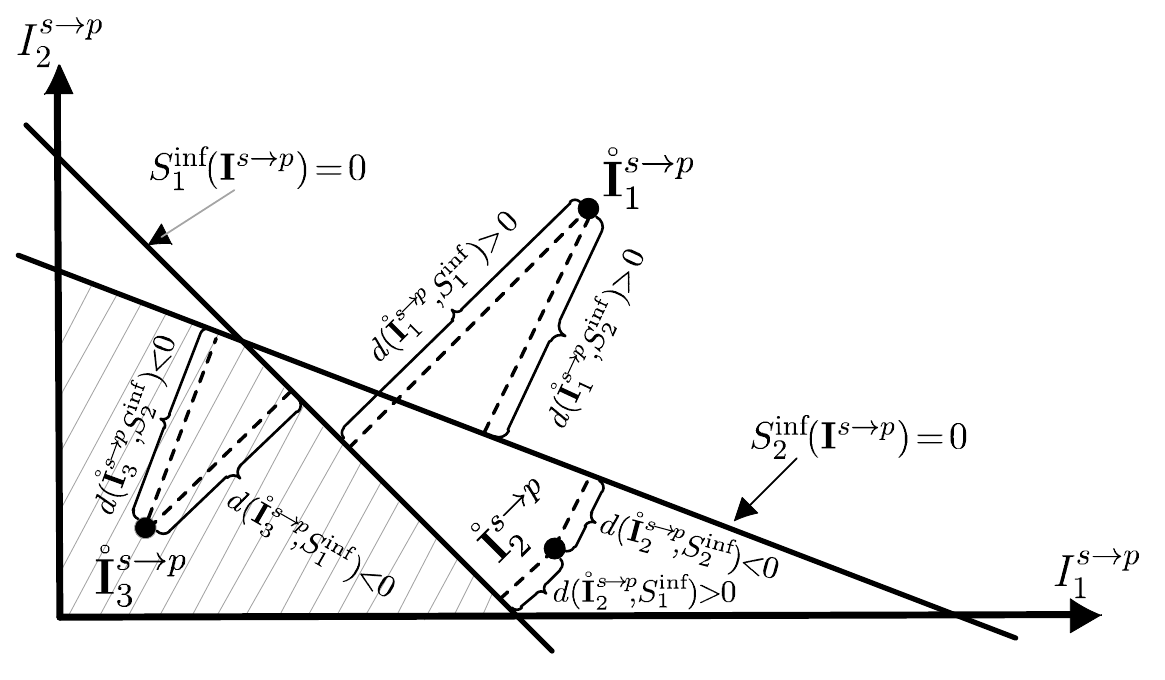} \\
				\vspace{-15pt}
		\caption{FCIR of a network with two PBSs showing the cognitive interference infeasibility boundaries $S_1^{\mathrm{inf}}\!(\mathbf{I}^{s\rightarrow p})\!=\!0$ and $S_2^{\mathrm{inf}}\!(\mathbf{I}^{s\rightarrow p})\!=\!0$ and the signed Euclidean distances of three possible cognitive interferences $\CIzerobold_1$, $\CIzerobold_2$, and $\CIzerobold_3$ from the infeasibility boundaries.} \vspace{-10pt}
		\label{fig:FCIR_example2}
	\end{figure*}

	\emph{Example 2:} Consider the CRN stated in Example 1. Fig. \ref{fig:FCIR_example2} shows the FCIR and the infeasibility  boundaries $S_1^{\mathrm{inf}}(\mathbf{I}^{s\rightarrow p})=0$ and $S_2^{\mathrm{inf}}(\CIbold)=0$  together with the infeasibility measures of three sample cognitive interference vectors $\CIzerobold_1$, $\CIzerobold_2$, and $\CIzerobold_3$ (marked as black-filled circles). For each given cognitive interference vector $\CIzerobold_m$, we have
	\begin{align*}
		d(\CIzerobold_m,S_1^{\mathrm{inf}})= 
		\frac{ A_{11}^p \CIzero_{m1} + A_{12}^p\CIzero_{m2} - C_1^{p}}
			{ \sqrt{\left(A_{11}^p\right)^2 + \left(A_{12}^p\right)^2}},
	\end{align*}
	and
	\begin{align*}
		d(\CIzerobold_{m},S_2^{\mathrm{inf}})= 
		\frac{ A_{21}^p\CIzero_{m1} + A_{22}^p \CIzero_{m2} - C_2^{p}}
			{ \sqrt{\left(A_{21}^p\right)^2 + \left(A_{22}^p\right)^2}}.
	\end{align*}
	It is seen that $\CIzerobold_1$ is infeasible for both of the PBSs, $\CIzerobold_2$ is only feasible for the second PBS, and $\CIzerobold_3$ is feasible for both of the PBSs.
	
	Given the transmit power vector $\mathring{\pbold}$, let $\CIzero_m(\A^s)$ denote the cognitive interference of the subset of SUs $\A^s\subset\M^s$ on the PBS $m\in\B^p$, i.e.,
	
	\begin{align}
	\label{eq:I_p_to_s_A}
		\CIzero_m(\A^s)=\sum_{i\in \A^s}{(\mathring{p}_i h_{mi})},\quad \forall m\in\B^p. 
	\end{align}
	We have
	\begin{align}
	\label{eq:I_p_to_s_A2}
		\CIzero_m(\A^s \! \setminus \! \{i\})= \CIzero_m(\A^s) - \mathring{p}_i h_{mi}, 
	\end{align}
	and therefore,
	\begin{align}
	\label{eq:I_p_to_s_A3}
		\CIzerobold(\A^s \! \setminus \! \{i\})= \CIzerobold(\A^s) - \mathring{p}_i\ [h_{1i},\ h_{2i},\ ...,\ h_{B^p i}]^{\mathrm{T}}. 
	\end{align}
	Based on the stated concept of the infeasibility measure, for the case where the stationary cognitive interference vector $\CIzerobold$ is infeasible for at-least one of the PBSs, we obtain the removal candidate SU (among active SUs) as the one whose removal results in the minimum aggregate infeasibility measure of the PBSs for which $\CIzerobold$ is infeasible,	i.e.,
	\begin{align}
	\label{eq:istarcase2}
		i^{*}=\argmin_{i\in \A^s} 
		\left\{ 
		{ \sum_{m\in B^p | S_m^{\mathrm{inf}}(\CIzerobold) > 0} 
			{ \hspace{-20pt}	d \left(\CIzerobold(\A^s \! \setminus \! \{i\}),S_m^{\mathrm{inf}}\right)} 
		}
		\right\}.
	\end{align}
	
	Based on what has been stated so far, our proposed JPAC algorithm can be stated as in \textbf{Algorithm 1}.
	\vspace{0.1cm}
	
	\startalg{\textbf{Algorithm 1:} JPAC for infeasible CRNs} 
		
	\begin{enumerate}[a)]
	\item \textit{Initialization} 
		
		\alg{ Assume all the SUs are admitted ($\A^s\leftarrow\M^s$).}
				
	\item \textit{Admission  Control}
		
		\newcommand{\ta}{\A^{h}_{\mathrm{srt}}(1)}
		\newcommand{\tb}{\A^{\thetahat}_{\mathrm{srt}}(1)}

		\alg{\textbf{While} $|\A^s|>0$ }
		\algstartblock	
			
			\alg{Obtain the power vector $\pboldzero$ in which $\pzero_i=0$ for each $i\in\M^s\setminus\A^s$, and $\pzero_i$ for the set of active users $\A^s\cup \M^p$ is obtained as the stationary power of constrained-TPC in \eqref{eq:constrained_TPC} .}
		
			\alg{Calculate $\CIzero_m(\A^s)$ for all $m\in\B^p$ from \eqref{eq:I_p_to_s_A}.}
		
			\alg{\textbf{Case 1:} $\CIzerobold(\A^s)$ lies within the FCIR (checked via \eqref{eq:322}) and there exists at least one SU $i\in\A^s$ for which $\gamma_i(\pboldzero)<\gammaihat$} 
			
			\algstartblock				
				
				\alg{Let \mbox{$\A^s\!\leftarrow\! \A^s \! \setminus \! \{i^{*}\}$}, where $i^{*}$ is obtained by \eqref{eq:23}.}
		
			\algendblock
			\alg{\textbf{Case 2:} $\CIzerobold(\A^s)$ is outside the FCIR} 
			\algstartblock				
				\alg{Let \mbox{$\A^s\!\leftarrow\! \A^s \! \setminus \! \{i^{*}\}$}, where $i^{*}$ is obtained by \eqref{eq:istarcase2}.}
			\algendblock
			\alg{\textbf{Case Else}} 
			\algstartblock				
				\alg{ \hspace{-8pt} \textbf{Exit While}}
			\algendblock
		\algendblock
		\alg{ \hspace{-6pt} \textbf{End}}
			
		\end{enumerate}
	\begin{remark}
		From Remark \ref{rmk:1}, as opposed to other JPAC algorithms which need all information pertinent to the PRN, our proposed algorithm can be run by the CRN with only feedback of the coefficients $\mathbf{A}^p$ and $\mathbf{C}^p$ (obtained from \eqref{eq:A_feas_coef} and \eqref{eq:B_feas_coef}, respectively) from the PRN. 
	\end{remark}
	
	\begin{remark}
		In order to determine the complexity of finding a removal candidate in our algorithm, we consider the worst case where there are maximum number of admitted SUs (i.e., when $|\A^s|=M^s$).
		The complexity of calculating either $m^{*}$ or $i^{*}$ in \eqref{eq:24} and \eqref{eq:23}, respectively, is of $O(M^s)$ and thus the overall complexity of finding $i^{*}$ in \eqref{eq:23} is of $O(M^s)$. On the other hand, the complexity of calculating $d(\CIzerobold,S_m^{\mathrm{inf}})$ from \eqref{eq:d} is of $O(B^p)$ and thus the complexity of calculating $i^{*}$ from \eqref{eq:istarcase2} is found to be of $O(B^p \times M^s)$. Therefore, the maximum order of complexity of finding the removal candidate in our algorithm is of $\max\{O(M^s),O(B^p \times M^s)\}=O(B^p \times M^s)$.
		
	\end{remark}
	
\subsection{Feasible System Case: Maximizing the Aggregate Throughput of SUs} 
	For a feasible system, the optimization problem of maximizing the aggregate throughput, as stated in \eqref{eq:opt_max_throughput}, is non-convex and may not be easily converted into a convex form.  We first write \eqref{eq:opt_max_throughput} as
	\begin{align}
	\label{eq:opt2_feas}
		\underset{\pbold^s}{\mathrm{maximize}}& \ \  \sum_{i\in\M^s}{ \log \bigg( 1+ \frac	{\hup{b_i}{i} p_i}
		 						{  \sum\limits_{\substack{j\in\M^s\\ j \ne i}}{\!\!\hup{b_i}{j} p_j} + \!
		 						I_{b_i}^{p \rightarrow s} +
		 						  \Nup{b_i} \!
		 						} \bigg) }  \ \ \   
	\nonumber \\
		\mathrm{subject\ to} 
		& \ \ \sum\limits_{n\in\B^p} {A^p_{mn} \sum_{i\in\M^s} {p_i h_{ni}} } -C^p_m \leq 0 , \ \forall m\in\B^p,
	\nonumber \\
		& \ \ \ 0\leq p_i \leq \pimax, \hspace{75pt} \forall i\in\M^s,
	\nonumber \\	
		& \ \ \gammaihat \leq \dfrac	{\hup{b_i}{i} p_i}
 						{  \sum\limits_{\substack{j\in\M^s\\ j \ne i}}{\!\!\hup{b_i}{j} p_j} + \!
 						I_{b_i}^{p \rightarrow s} +
 						  \Nup{b_i} \!
 						}, \hspace{10pt} \forall i\in\M^s,
	\end{align}
	where the first constraint corresponds to the protection of PUs and the second and third constraints correspond to the protection of SUs and $I_{b_i}^{p \rightarrow s}$ 
	is the interference caused by all the PUs to the receiving-point of SU $i\in\M^s$. 
	The problem formulation of \eqref{eq:opt2_feas} is similar to the maximum throughput optimization problems considered in many of the existing works except that the first constraint for PUs' protection is modified by the constraint corresponding to the characterized polyhedron FCIR. It can be easily verified that \eqref{eq:opt2_feas} is equivalent to the following optimization problem:
	\begin{align}
	\label{eq:opt3_feas}
		\underset{\pbold^s,\gammabold^s}{\mathrm{maximize}}& \ \ \log \prod_{i\in\M^s}{  ( 1+ \gamma_i ) }  \ \ \   
	\nonumber \\
		\mathrm{subject\ to} 
		& \ \ \sum\limits_{n\in\B^p} { \frac{A^p_{mn}}{C^p_m} \sum_{i\in\M^s} {p_i h_{ni}} } \leq 1, &\forall m\in\B^p,
	\nonumber \\
		& \ \  p_i (p_i^{\mathrm{max}})^{-1} \leq 1, & \forall i\in\M^s, 
	\nonumber \\
		& \ \ p_i \geq 0, \quad \forall i\in\M^s, 
	\nonumber \\
		& \ \ \gammahat_i \gamma_i^{-1}  \leq 1, \quad & \forall i\in\M^s, 
	\nonumber \\	
		& \ \ \gamma_i
		\frac{\sum\limits_{\substack{j\in\M^s, j \ne i}}{\!\!\hup{b_i}{j} p_j} + \!
				 						I_{b_i}^{p \rightarrow s} +
				 						  \Nup{b_i}}
		{\hup{b_i}{i} p_i}
		\leq 1
 						, & \forall i\in\M^s. 
	\end{align}
	By using \textit{Successive Geometric Programming} (\cite{power_control_by_GP,successive_inner_approximation}) and taking similar steps to the ones in \cite{GP_bad_ITL}, we can revise the maximum throughput power control algorithm proposed in \cite{GP_bad_ITL} as the  algorithm stated in \textbf{Algorithm 2}.

	\startalgg{\textbf{Algorithm 2:} Successive geometric program to find the maximum aggregate throughput of  SUs in a feasible system}
	
	\begin{enumerate}[a)]
		\item \textit{Initialization} 
		\alg{For all $i\in\M^s$, initialize ${\gamma}_i^0\geq 0$ and ${p}_i^0 \geq 0$.}
	\item \textit{Successive inner approximation}
		
		\newcommand{\ta}{\A^{h}_{\mathrm{srt}}(1)}
		\newcommand{\tb}{\A^{\thetahat}_{\mathrm{srt}}(1)}

		\alg{\textbf{Do}}
		\algstartblock	
			\alg{Let $\lambda_i=\frac{\gamma_i}{\gamma_i + 1},\forall i\in\M^{s}$ and $c=\prod_{i\in\M^s}{\frac{(1 + \gamma_i)}{(\gamma_i)^{\lambda_i}}}$.}
			\alg{Solve the following geometric program:} 
				\begin{align}
				 \underset{\widetilde{\pbold}^s,\widetilde{\gammabold}^s}{\mathrm{maximize}} \ c \! \prod_{i\in\M^s}{  {\!\! \gamma_i}^{\lambda_i} }  \   
					 \mathrm{s.t.}  
					 \ \textrm{constraints of\ } \eqref{eq:opt3_feas}.	\nonumber					
				\end{align}
		\algendblock
		
		\alg{ \textbf{Loop Until} the algorithm converges.}
	
	
			
		\end{enumerate}	
		

	
\section{Numerical Results}
		\label{sec:numerical_results}
	Consider a wireless network consisting of a PRN and a CRN. Consider the noise power levels at all receiving-points to be $5\times{10}^{-13}$ Watts. We consider the following model for path-gain: $h_{b_i j}=k d_{ij}^{-4}$, where $d_{ij}$ is the distance between user $j$ and the corresponding receiver of user $i$ and $k$ is the attenuation factor. We assume $k=0.09$ and $\pimax=0.1$ Watts for all users.

\subsection{Illustration of FCIR for Different Network Scenarios}

\begin{figure}
		\centering
		\includegraphics [width=180pt,height=200pt]{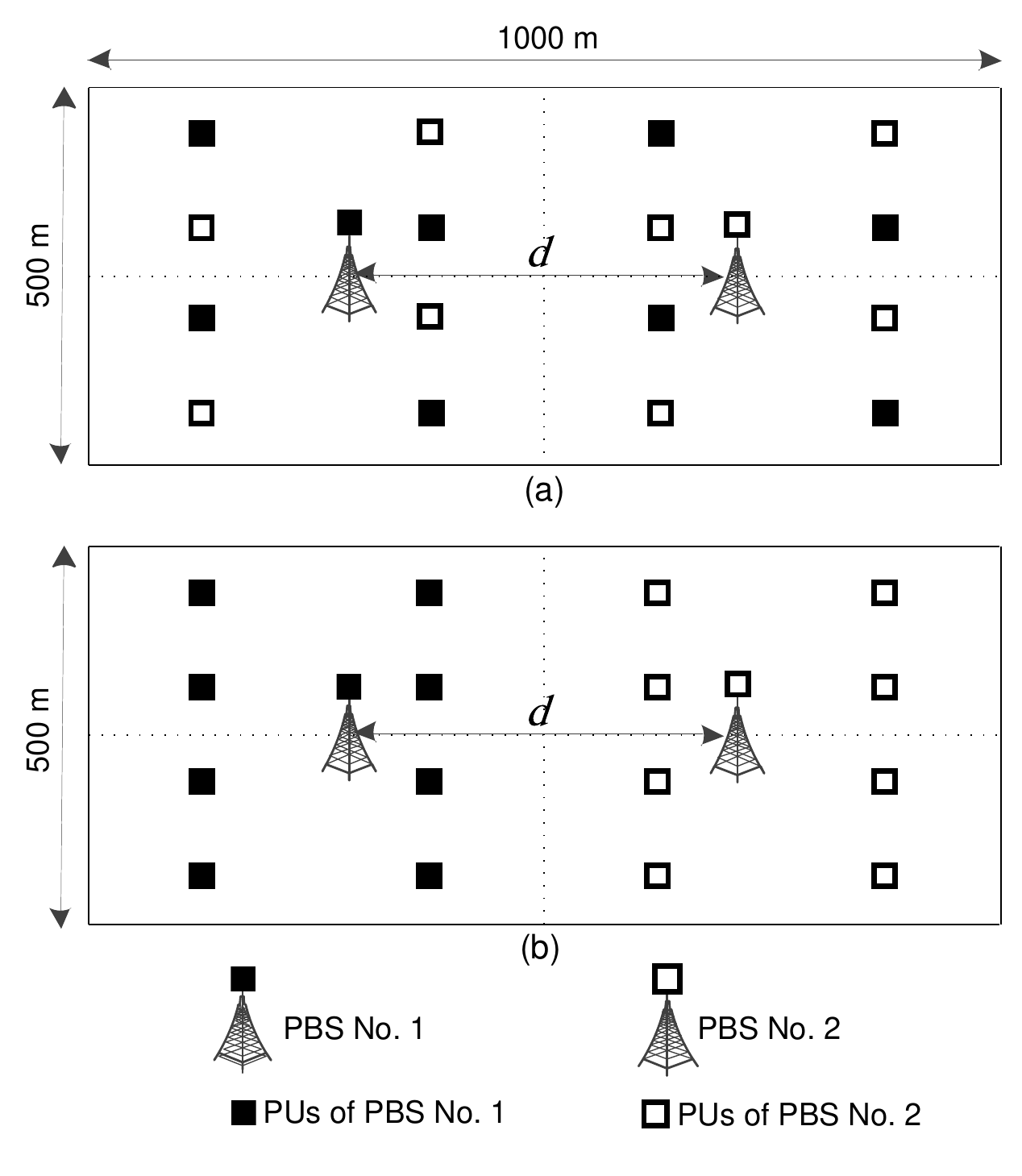}\\
		\caption{Two scenarios are considered to show the FCIR. In scenario (a),  all the PUs are evenly distributed throughout the whole PRN, and in scenario (b), the PUs are evenly distributed in the area close to their corresponding BSs.} \vspace{-10pt}
		\label{fig:sim_main_topology}
	\end{figure}

	First we present numerical results to show how the FCIR varies for different primary network scenarios. Consider that PUs are located in a \mbox{500 m $\times$ 1000 m} area covered by two PBSs located at the points $(500-d/2,250)$ and $(500+d/2,250)$, respectively, both having the height of $20$ m (see Fig. \ref{fig:sim_main_topology}). Since the SUs' parameters do not affect the FCIR, we do not depict the CRN and SUs in Fig. \ref{fig:sim_main_topology}. We consider two different scenarios. For the first scenario, all of the PUs are evenly located throughout the PRN area and assigned to the BSs as in Fig. \ref{fig:sim_main_topology}(a). For the second scenario, all of the PUs are evenly spread in the PRN and assigned to the nearer PBS as in Fig. \ref{fig:sim_main_topology}(b).

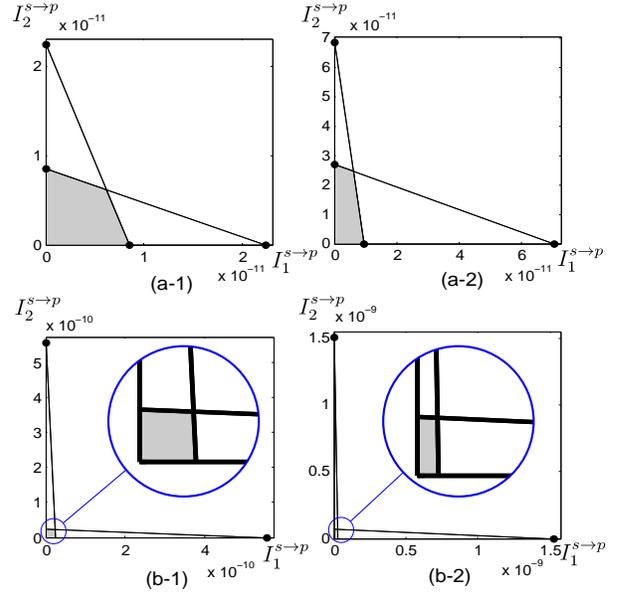
\begin{figure}
		\begin{tikzpicture}[spy using outlines=
			{circle, magnification=6, connect spies}]
			\begin{axis}[hide axis,enlargelimits=false,axis on top,axis equal image,
				width=320pt,
				xmin=0,ymin=0,xmax=1,ymax=1,
			]
				\addplot graphics [xmin=0,ymin=0,xmax=1,ymax=1] {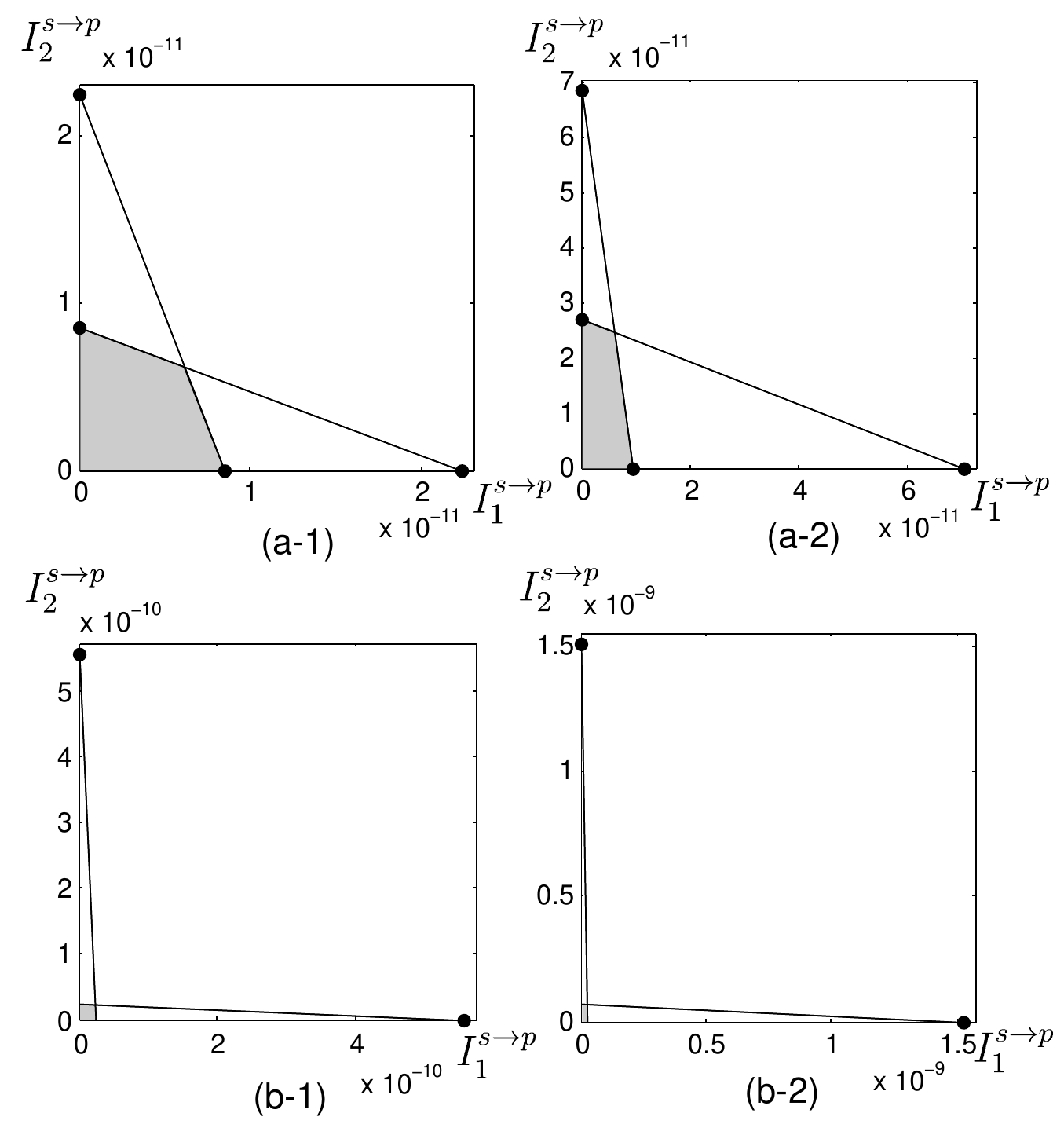};
				\coordinate (spypoint) at (axis cs:0.087,0.120);
		  		\coordinate (magnifyglass) at (axis cs:0.3,0.3);
		  		\coordinate (spypoint2) at (axis cs:0.558,0.122);
				\coordinate (magnifyglass2) at (axis cs:0.75,0.3);
			\end{axis}
			\spy [blue, size=2cm] on (spypoint) in node[fill=white] at (magnifyglass);
			\spy [blue, size=2cm] on (spypoint2) in node[fill=white] at (magnifyglass2);
		\end{tikzpicture}
	\caption{Different regions of the FCIR; (a-1) and (a-2) for the case where the PUs are located according to Fig. \ref{fig:sim_main_topology}(a) and (b-1) and (b-2) for the case where the PUs are located according to Fig. \ref{fig:sim_main_topology}(b). In (a-1) and (b-1), all the PUs have the target-SINR of $-18$ dB. In (a-2) and (b-2) all PUs associated with PBS no. 1 and all PUs associated with PBS no. 2 have the target-SINR of $-18$ and $-22$ dB, respectively.} \vspace{-10pt}
	\label{fig:sim_FCIR_regions}
	\end{figure}

	The FCIR corresponding to different network scenarios are shown in Fig. \ref{fig:sim_FCIR_regions}. Figs. \ref{fig:sim_FCIR_regions}(a-1) and \ref{fig:sim_FCIR_regions}(a-2) are for the case where the PUs are located according to Fig. \ref{fig:sim_main_topology}(a). Figs. \ref{fig:sim_FCIR_regions}(b-1) and \ref{fig:sim_FCIR_regions}(b-2) are for the case where the PUs are located according to Fig. \ref{fig:sim_main_topology}(b). Besides, Figs. \ref{fig:sim_FCIR_regions}(a-1) and \ref{fig:sim_FCIR_regions}(b-1) show the FCIR for the case where all PUs have the same SINR of $-18$ dB and Figs. \ref{fig:sim_FCIR_regions}(a-2) and \ref{fig:sim_FCIR_regions}(b-2) show the FCIR for the case where the PUs assigned with the first and second PBSs have the SINR of $-18$ and $-22$ dB, respectively.

	It can be observed from Figs. \ref{fig:sim_FCIR_regions}(a-1) and \ref{fig:sim_FCIR_regions}(b-1) that the symmetric distribution of the PUs and the same conditions of the two cells of the PRN lead to a symmetric value-region of the FCIR. On the other hand, it is seen from Figs. \ref{fig:sim_FCIR_regions}(a-2) and \ref{fig:sim_FCIR_regions}(b-2) that since the PUs associated with the second PBS have lower SINR targets, the second PBS can tolerate more cognitive interference as compared with that of the first PBS. 
	
\subsection{Performance Evaluation of the Proposed JPAC Algorithm for Infeasible Systems}
\begin{figure*}
		\centering
		\includegraphics []{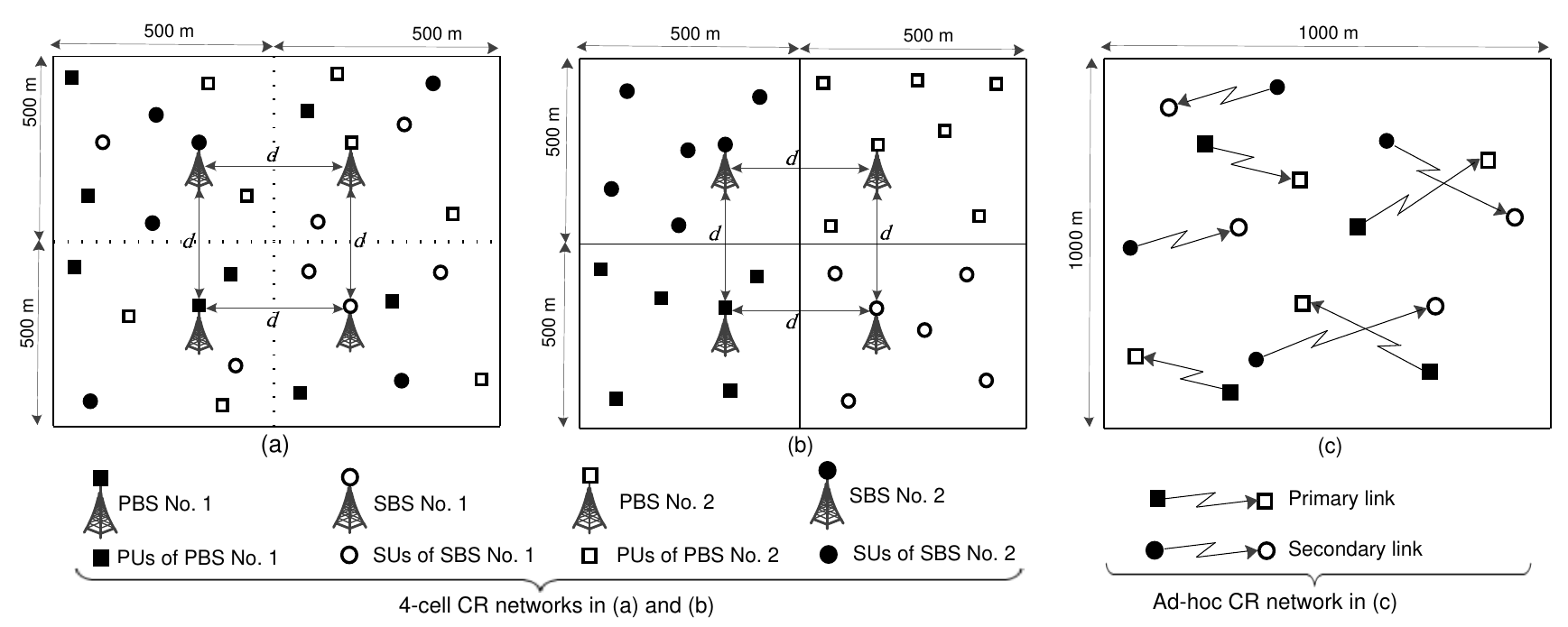}\\
		\caption{Three different scenarios for the evaluation of the performance of our proposed algorithms:  (a) and (b) show 4-cell networks where the users are randomly spread throughout the network area and in the area closer to their serving BSs respectively, and (c) shows an ad-hoc network where the transmitter-receiver pairs are randomly spread in the network area.} \vspace{-10pt}
		\label{fig:sim_Topology3}
	\end{figure*}

	To show the performance of our proposed JPAC algorithm and demonstrate the impact of considering fixed  ITL values for the PBSs on the performance of existing algorithms, we consider a $1000\times 1000$ m area for three different scenarios. In the first two scenarios, we consider two 4-cell  networks wherein two PBSs and two SBSs having the height of 20 m are located in $(\pm \frac{d}{2}, \pm \frac{d}{2})$ according to Figs. \ref{fig:sim_Topology3}(a) and \ref{fig:sim_Topology3}(b). In Fig. \ref{fig:sim_Topology3}(a), it is assumed that all the PUs and SUs are randomly spread throughout the network area. In Fig. \ref{fig:sim_Topology3}(b), it is assumed that the PUs and SUs of each cell are spread in the rectangular area closer to their serving BSs as compared to other BSs. In the third scenario, we consider an ad-hoc network as shown in Fig. \ref{fig:sim_Topology3}(c), where the primary and secondary links (transmitter-receiver pairs) are randomly located in the coverage area of the network where the receiving-point of each link has the maximum distance of 250 m from the transmitter of the corresponding link.

	In all simulation scenarios, we consider that each user's (i.e., PU's or SU's) target-SINR is randomly chosen from the set of $\{-20, -24\}$ dB for the 4-cell network as in Fig. \ref{fig:sim_Topology3}(a), $\{-12, -16\}$ dB for the 4-cell network as in Fig. \ref{fig:sim_Topology3}(b) and $\{-16, -20\}$ dB for the ad-hoc network as in Fig. \ref{fig:sim_Topology3}(c). To evaluate the performance of our proposed algorithm, we compare our proposed Algorithm 1 with the well-known JPAC algorithms of ISMIRA (\cite{ISMIRA}) and LGRA (\cite{LGRA}). However, as has been stated before, these related algorithms  consider fixed ITL values for the PBSs. 
	To calculate such values, for each PBS $m\in\B^p$, we  define $\overline{I}^p_{0,m}$ to be the maximum tolerable cognitive interference of the corresponding PBS, when the cognitive interference caused to all other PBSs are zero. From \eqref{eq:322} and also as seen in Fig. \eqref{fig:ex_FTIR_FCIR}, $\overline{I}^p_{0,m}$ is obtained as 
	\begin{align}
	\label{eq:232}
		\overline{I}^p_{0,m}=\min\limits_{n\in\B^p} {\frac{C_n^p}{A^p_{nm}}}.
	\end{align}
	The ITL for each PBS $i\in\B^p$ needed for ISMIRA and LGRA is obtained as
	\begin{align}
		\label{eq:233}
		\overline{I}^p_{m}=\alpha \overline{I}^p_{0,m},
	\end{align}
	where $\alpha$ is a constant. To compare the performance of our proposed \textbf{Algorithm 1} with that of the ISMIRA and LGRA, we consider different values of $\alpha$ for each of the ISMIRA and LGRA algorithms. Note that the protection of PUs is guaranteed for $\alpha\leq \alpha_0$, where $\alpha_0$ is some constant value lower than unity. All of the performance data presented in the following are obtained by averaging the results from 3000 independent snapshots.  
	
\subsubsection{Performance under varying number of SUs}
	
\begin{figure}
		\centering
		\includegraphics [width=254pt,height=110pt]{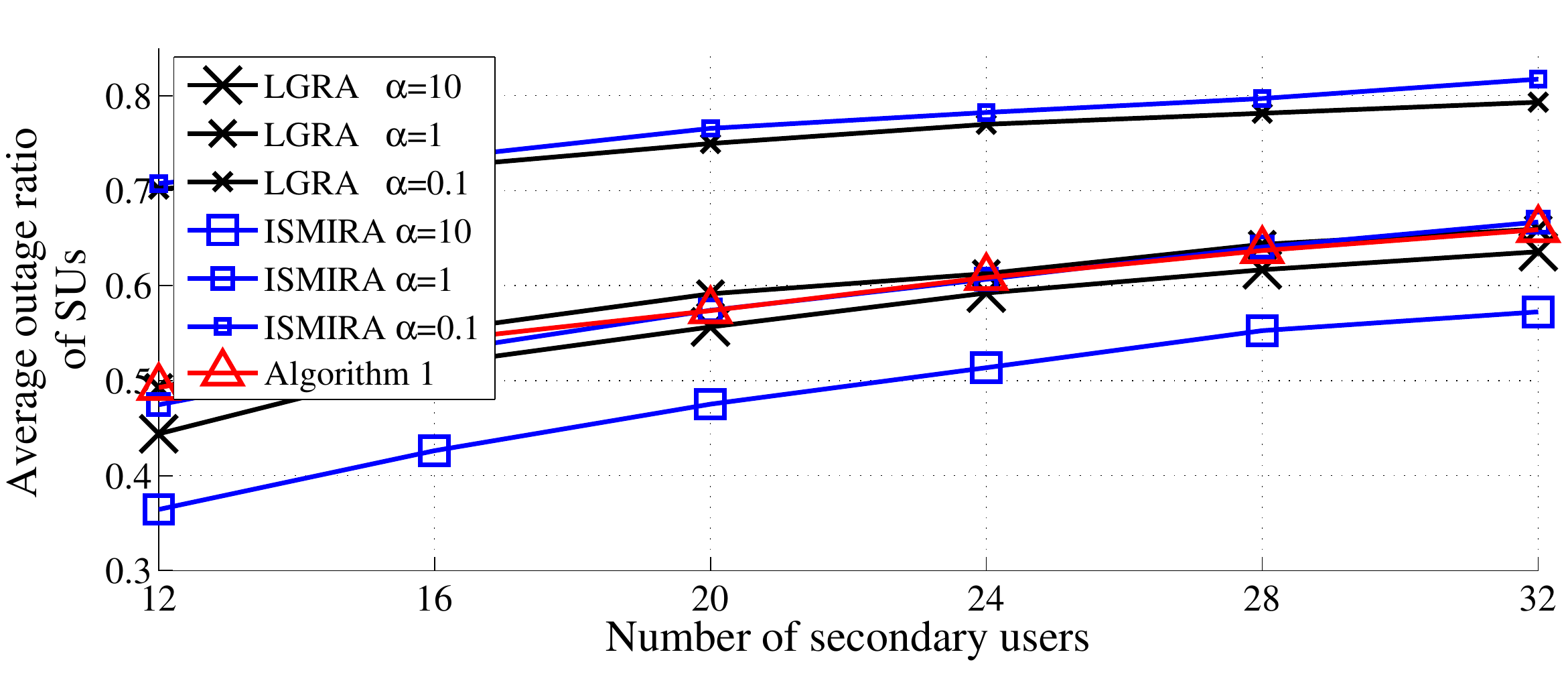}\\ 
		\includegraphics [width=254pt,height=110pt]{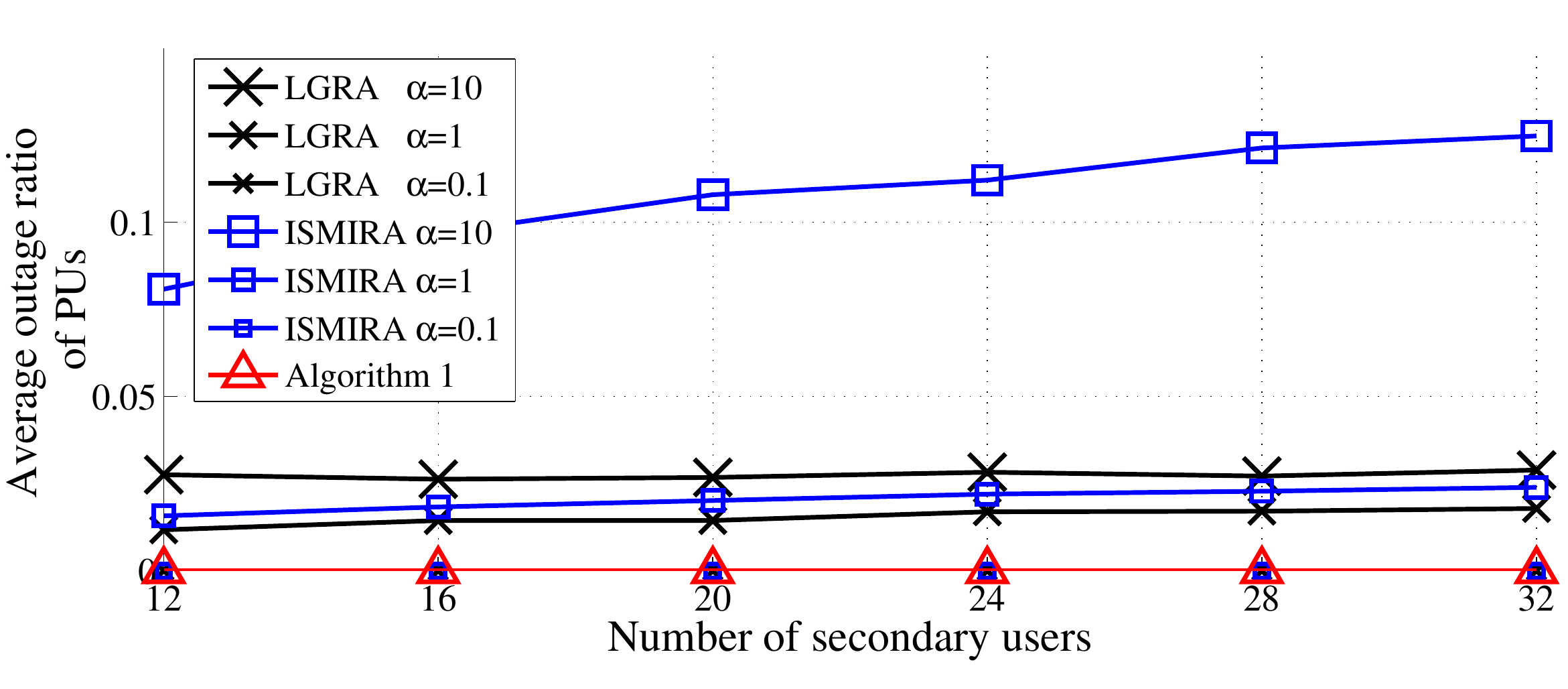}\\ %
		\caption{Average outage ratios of SUs and PUs for Algorithm 1 and for ISMIRA and LGRA with $\alpha=0.1$, $\alpha=1$, and $\alpha=10$ versus different total number of SUs for the  4-cell network scenario according to Fig. \ref{fig:sim_Topology3}(a).}
	\label{fig:sim_versus_users_1}
	\end{figure}

\begin{figure}
		\centering
		\includegraphics [width=254pt,height=110pt]{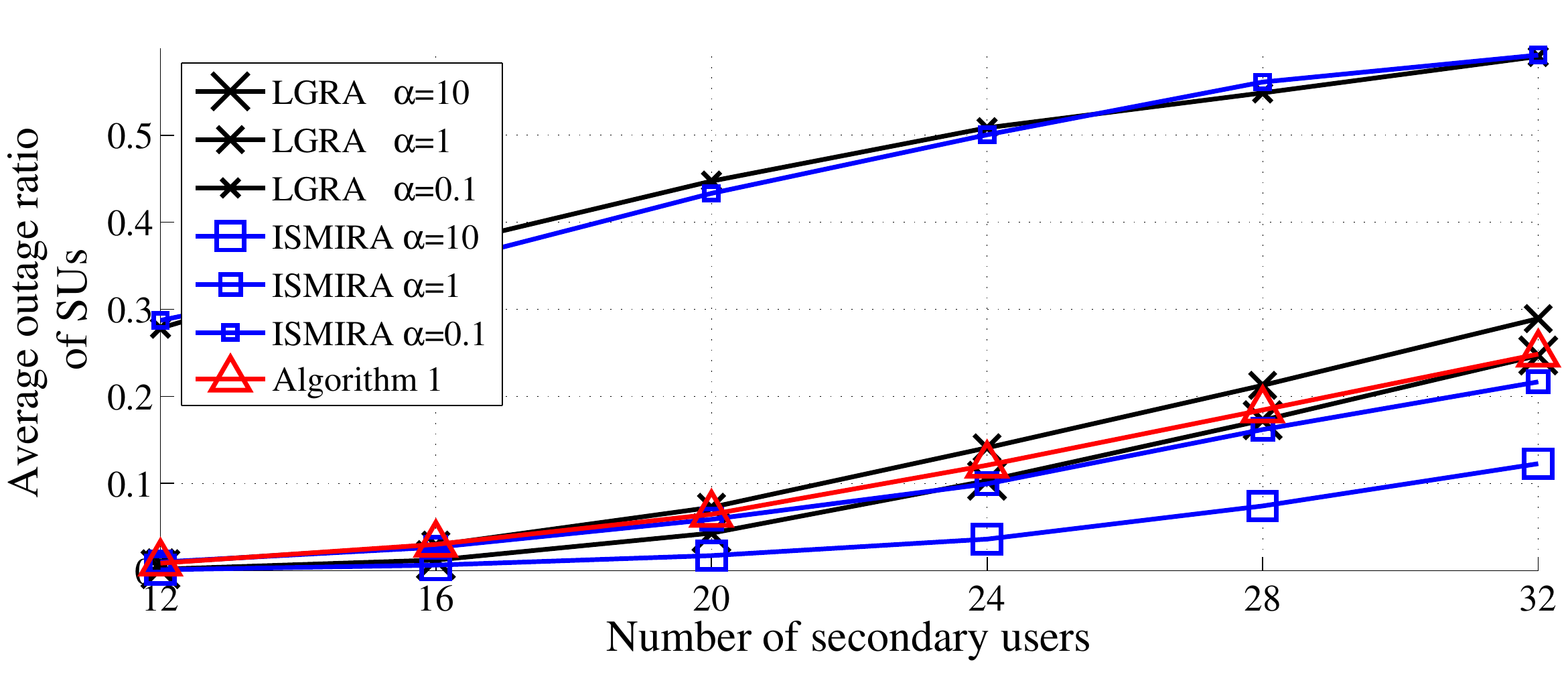}\\ 
		\includegraphics [width=254pt,height=110pt]{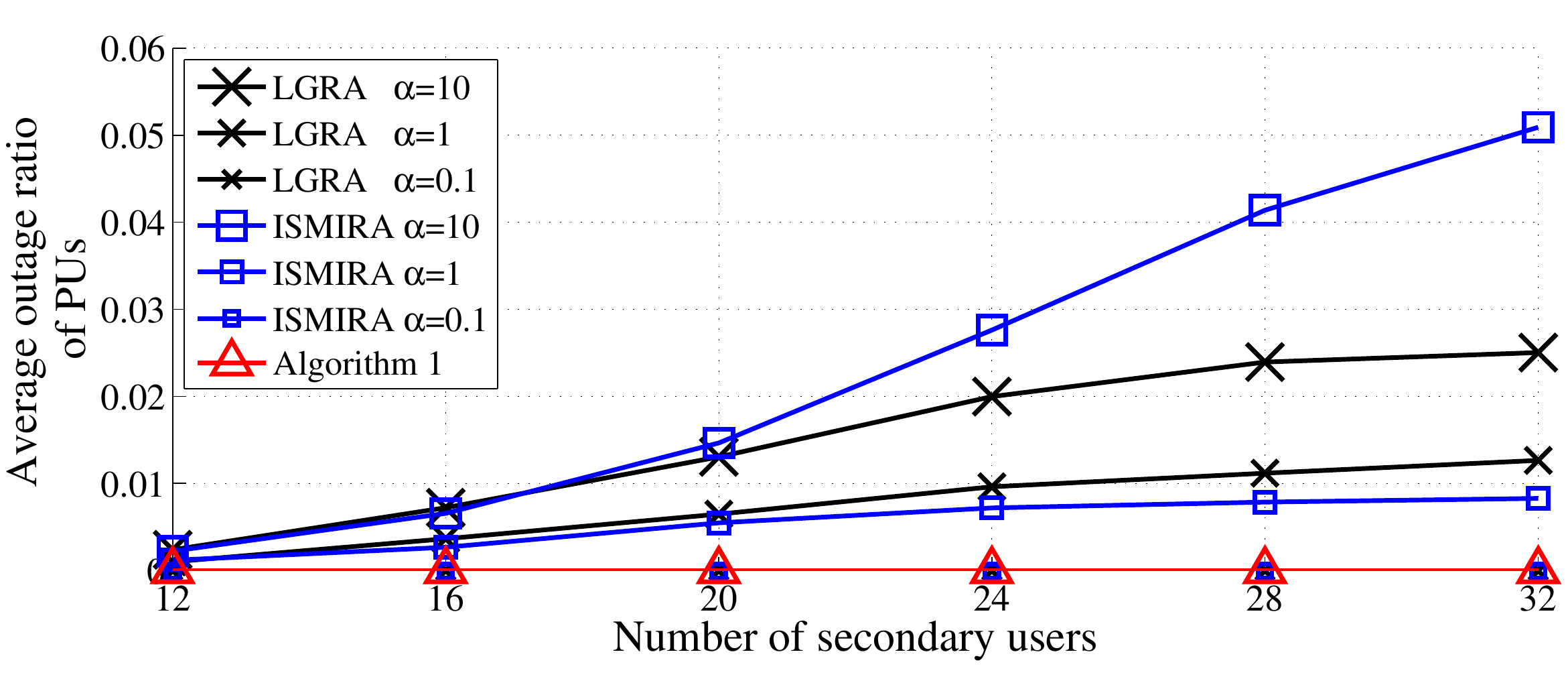}\\ %
		\caption{Average outage ratios of SUs and PUs for Algorithm 1 and for ISMIRA and LGRA with $\alpha=0.1$, $\alpha=1$, and $\alpha=10$ versus different total number of SUs for  the 4-cell network scenario according to Fig. \ref{fig:sim_Topology3}(b).}
	\label{fig:sim_versus_users_2}
	\end{figure}

	Consider the case where $20$ PUs are randomly located in the coverage area of the first and second PBSs and different number of SUs (varying from 12 to 32 with the step size of 4 SUs) are randomly located in the coverage area of the first and second SBSs in the 4-cell networks of Fig. \ref{fig:sim_Topology3}(a) and \ref{fig:sim_Topology3}(b). We assume \mbox{$d=150$ m}. Figs. \ref{fig:sim_versus_users_1} and \ref{fig:sim_versus_users_2} show the average outage ratios of the SUs and PUs for different total number of SUs, for the 4-cell scenarios in Figs. \ref{fig:sim_Topology3}(a) and \ref{fig:sim_Topology3}(b), respectively, for our proposed Algorithm 1 in comparison  with ISMIRA and LGRA for three different values of $\alpha$ ($\alpha=0.1$, $\alpha=1$, and $\alpha=10$). It is seen that the average outage ratios for all the algorithms increase as the total number of SUs increases and the average outage ratio obtained for \textbf{Algorithm 1} is lower than that for ISMIRA and LGRA for $\alpha=0.1$. While the protection of the PUs is always guaranteed for \textbf{Algorithm 1}, it is seen from Figs. \ref{fig:sim_versus_users_1} and \ref{fig:sim_versus_users_2} that the protection of PUs is violated for $\alpha=1$ and $\alpha=10$ for each of the ISMIRA and LGRA algorithms. That is, these algorithms do not guarantee the protection of PUs for such values of $\alpha$ (because they have been designed based on the assumption of fixed values for ITLs).
	
\subsubsection{Performance under varying distance between BSs}
\begin{figure}
		\centering
		\includegraphics [width=254pt,height=110pt]{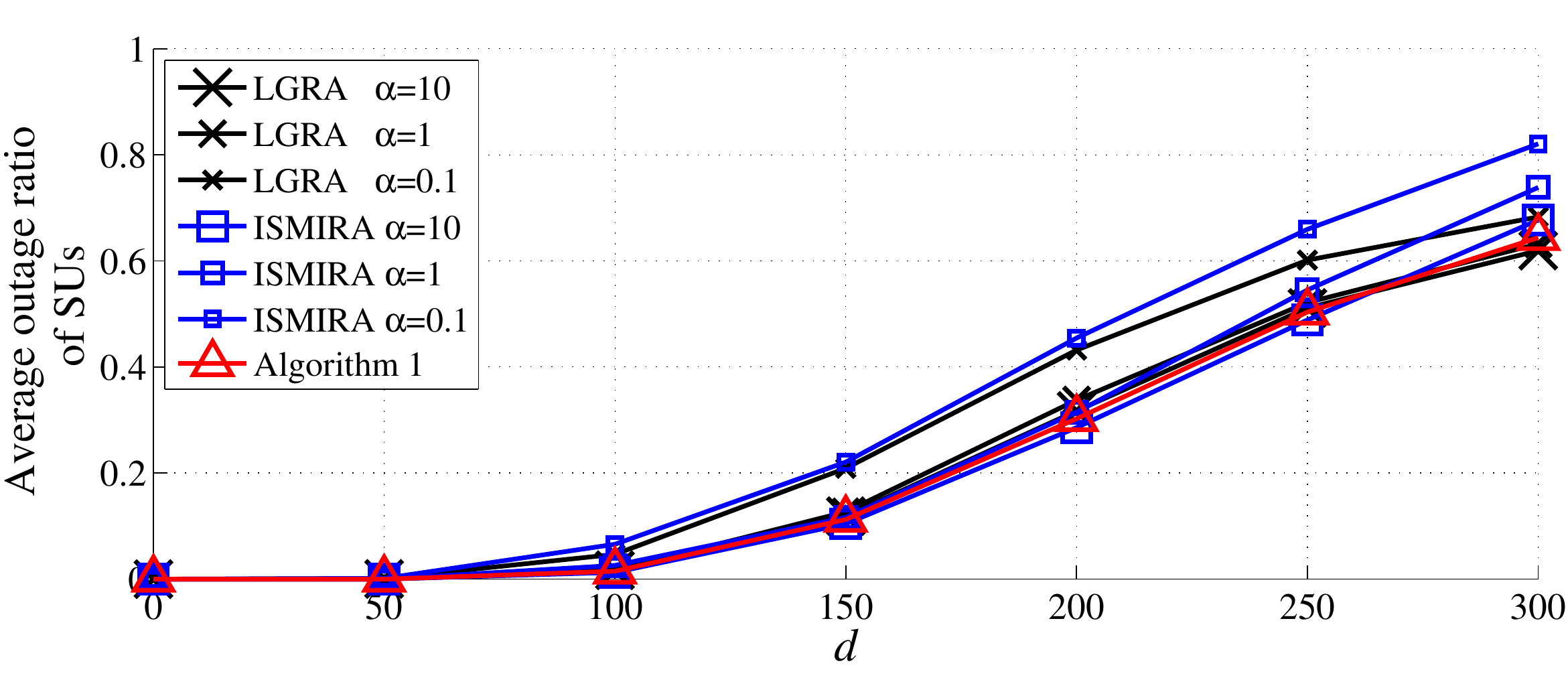}\\ 
		\includegraphics [width=254pt,height=110pt]{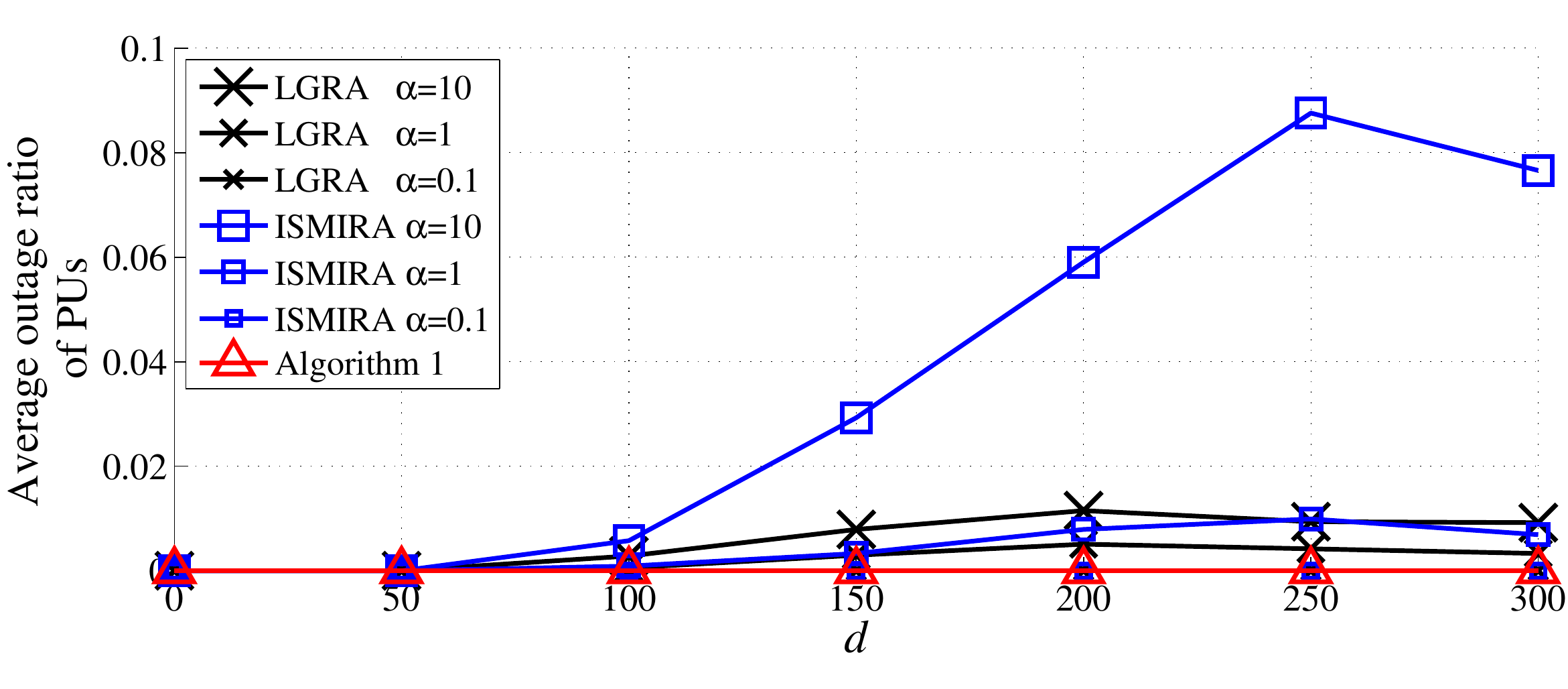}\\ %
		\caption{Average outage ratios of SUs and PUs for Algorithm 1 and for ISMIRA and LGRA with $\alpha=0.1$, $\alpha=1$, and $\alpha=10$ versus different values of $d$ for the 4-cell network scenario according to Fig. \ref{fig:sim_Topology3}(a).}
	\label{fig:sim_outage_versus_distance_of_BS1}
	\end{figure}

\begin{figure}
		\centering
		\includegraphics [width=254pt,height=110pt]{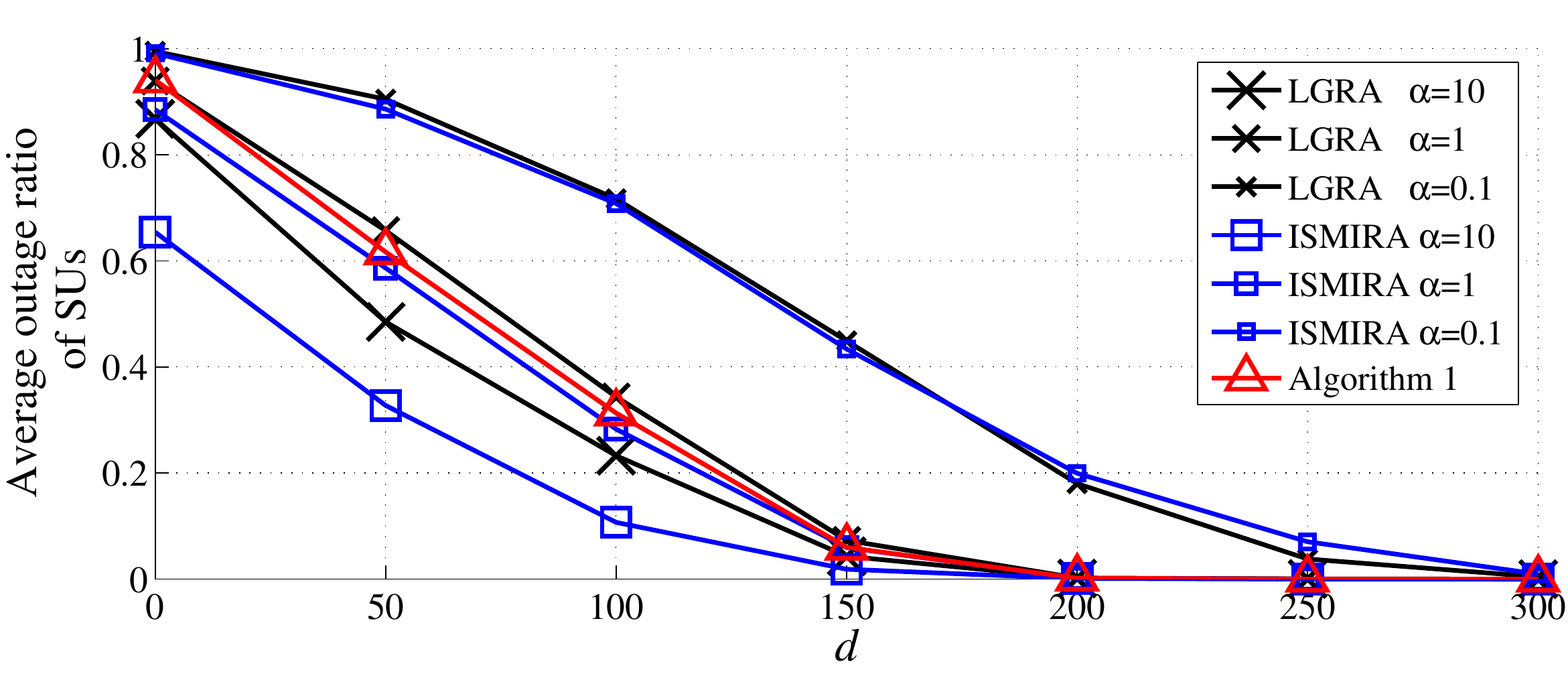}\\ 
		\includegraphics [width=254pt,height=110pt]{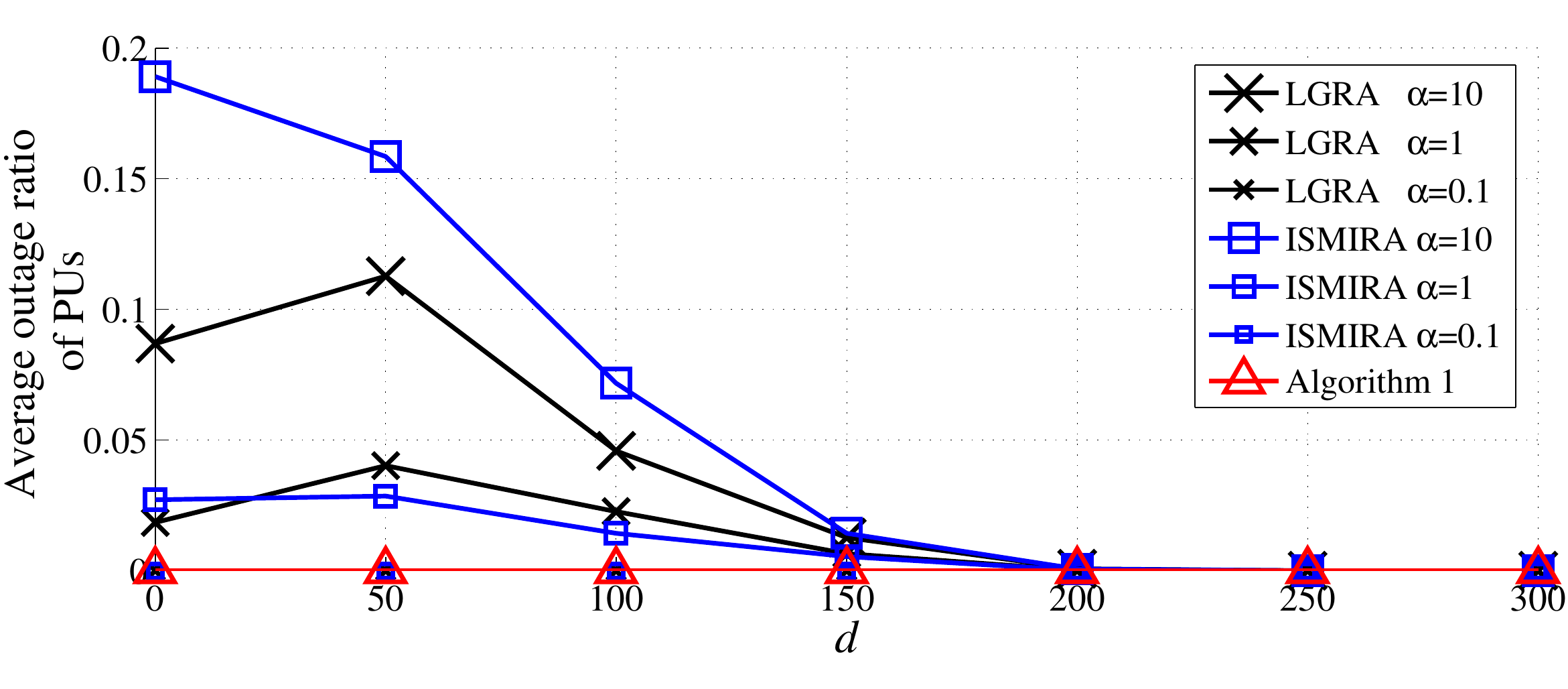}\\ %
		\caption{Average outage ratios of SUs and PUs for Algorithm 1 and for ISMIRA and LGRA with $\alpha=0.1$, $\alpha=1$, and $\alpha=10$ versus different values of $d$ for  4-cell network scenario according to Fig. \ref{fig:sim_Topology3}(b).}
	\label{fig:sim_outage_versus_distance_of_BS2}
	\end{figure}

	To show how the distance between BSs (denoted by $d$ in Figs. \ref{fig:sim_Topology3}(a) and \ref{fig:sim_Topology3}(b)) affects the performance of the algorithms, consider the case where $20$  PUs and $20$  SUs are randomly located in the coverage area of their corresponding BSs in any of the 4-cell networks as in Fig. \ref{fig:sim_Topology3}(a) and \ref{fig:sim_Topology3}(b). We consider different values of $d$ ranging from 0 to 300 m with the step-size of 50 m.
	Figs. \ref{fig:sim_outage_versus_distance_of_BS1} and \ref{fig:sim_outage_versus_distance_of_BS2} show the average outage ratios of the SUs and PUs for different values of $d$ under the 4-cell scenarios depicted in Figs. \ref{fig:sim_Topology3}(a) and \ref{fig:sim_Topology3}(b), respectively, for our proposed \textbf{Algorithm 1} in comparison with ISMIRA and LGRA for $\alpha=0.1$, $\alpha=1$ and $\alpha=10$. It is seen that the average outage ratio of the SUs for \textbf{Algorithm 1} is always lower than that of ISMIRA and LGRA for $\alpha=0.1$ for both scenarios. Similar to the results of the previous subsection, while the SUs' average outage ratio in ISMIRA and LGRA for $\alpha=10$ is seen to be smaller than that of \textbf{Algorithm 1}, it is seen that the constraint of protecting the PUs is violated for these algorithms for $\alpha=1$ and $\alpha=10$ which makes the comparison for SUs' protection nonsense. 
	It is also seen that, in Fig. \ref{fig:sim_outage_versus_distance_of_BS1} where users are spread according to Fig. \ref{fig:sim_Topology3}(a), for each SBS, an increase in $d$ results in an increase in the average distance between that SBS and its SUs leading to a higher average outage ratio for the SUs. On the other hand, for Fig. \ref{fig:sim_outage_versus_distance_of_BS2} where the users are spread according to Fig. \ref{fig:sim_Topology3}(b), for each SBS, an increase in $d$ results in an increase in the average distance between that SBS and other cells' interfering users, leading to a lower average outage ratio for the SUs.
	
\subsubsection{Performance of \textbf{Algorithm 1} Compared to Existing Algorithms Under the Assumption of Box-like FCIR}
\begin{figure}
		\centering
		\includegraphics [width=254pt,height=110pt]{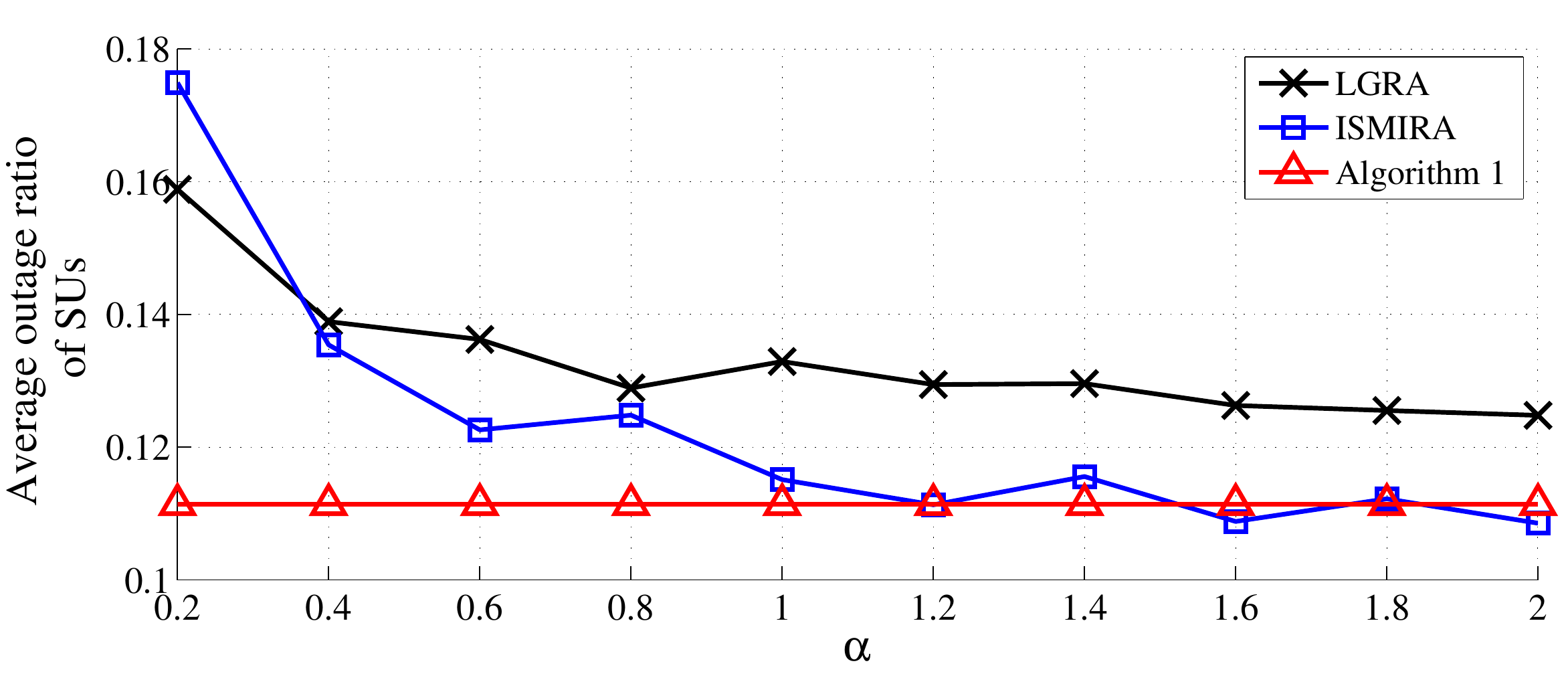}\\ 
		\includegraphics [width=254pt,height=110pt]{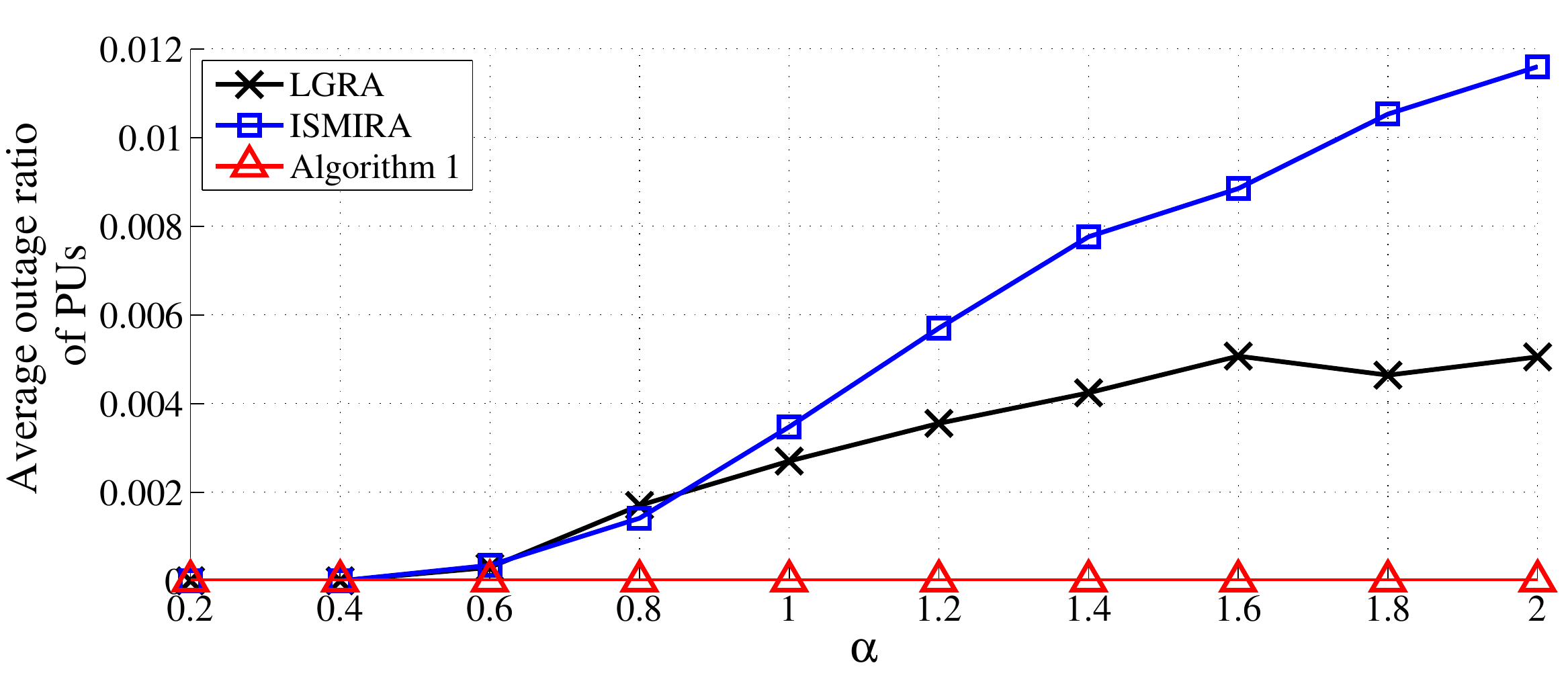}\\ %
		\caption{Average outage ratios of SUs and PUs for Algorithm 1 and for ISMIRA and LGRA with box-like FCIR versus varying box size in terms of $\alpha$ for the 4-cell network scenario according to Fig. \ref{fig:sim_Topology3}(a).}
	\label{fig:sim_outage_versus_alpha_4cells1}
	\end{figure}

\begin{figure}
		\centering
		\includegraphics [width=254pt,height=110pt]{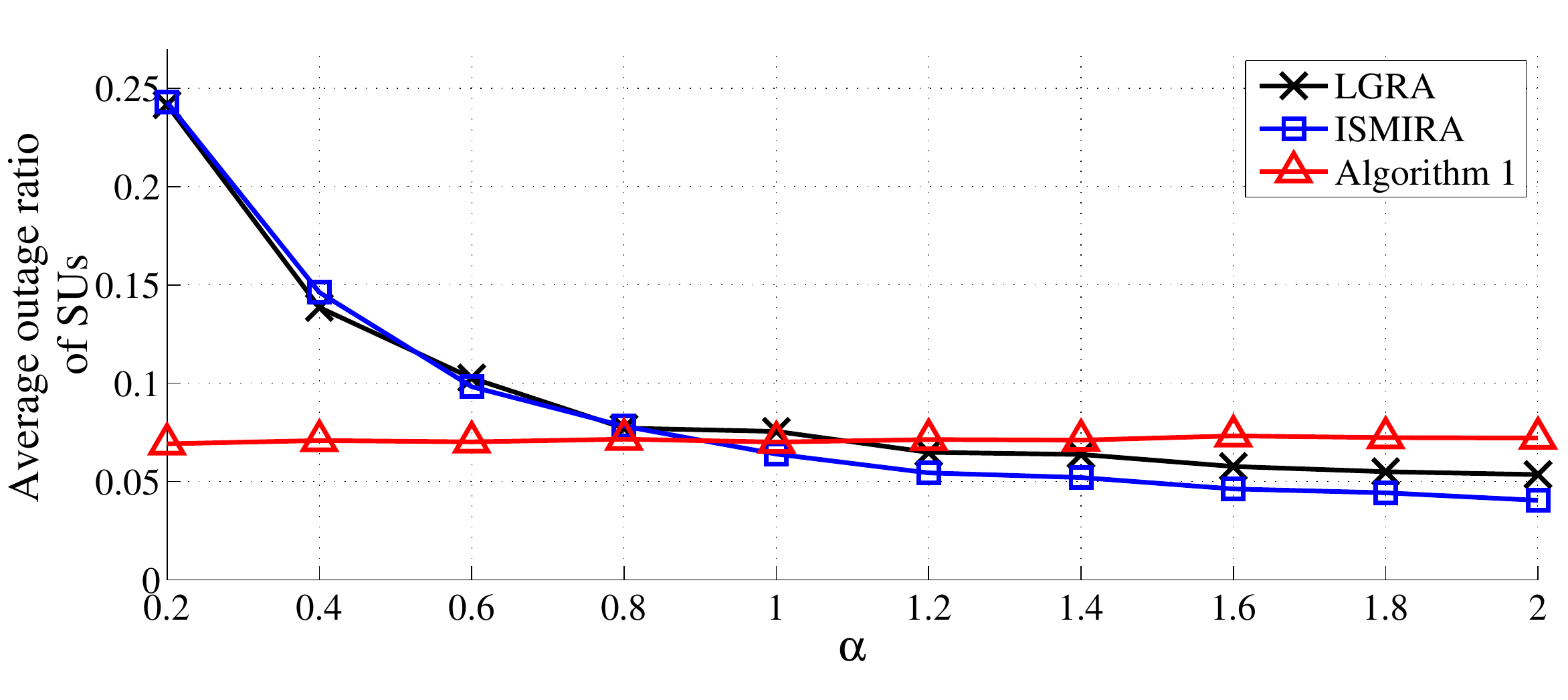}\\ 
		\includegraphics [width=254pt,height=110pt]{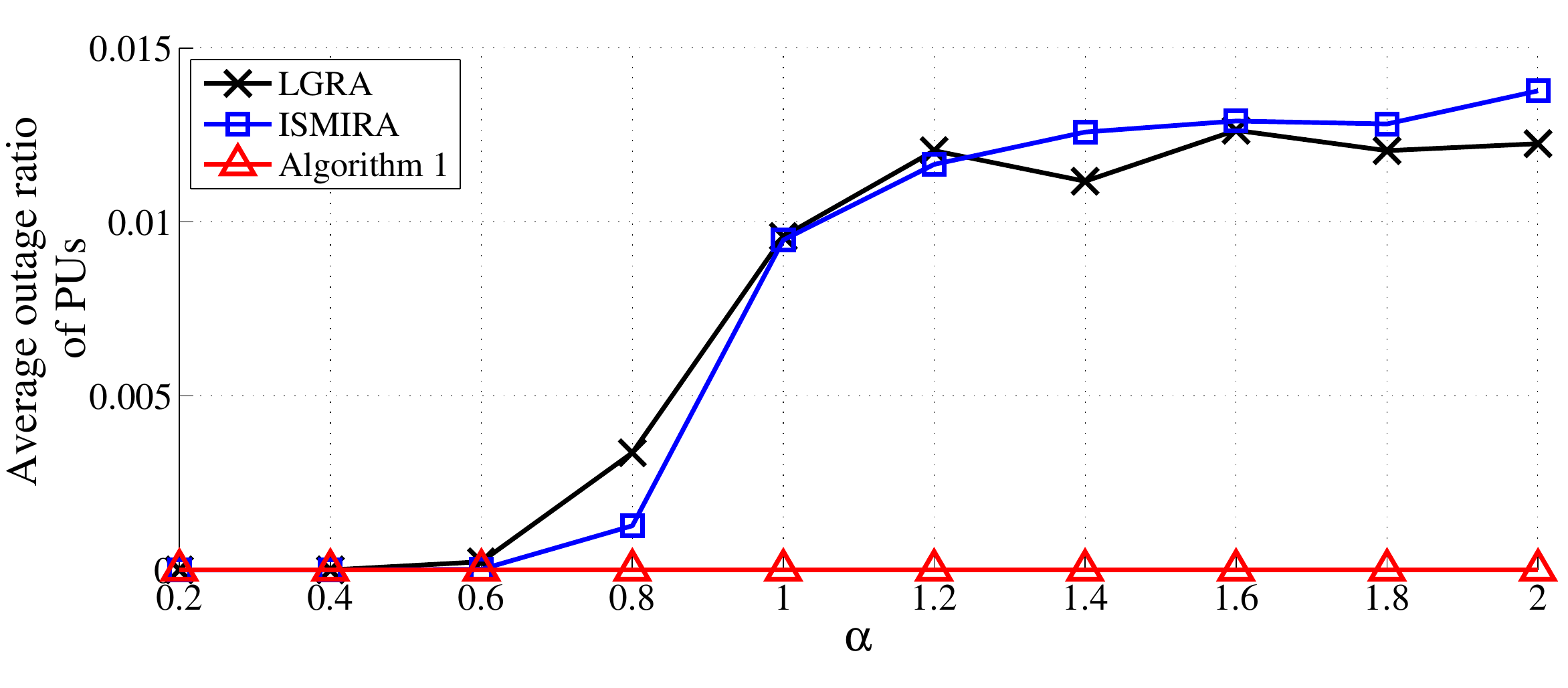}\\ %
		\caption{Average outage ratios of SUs and PUs for Algorithm 1 and for ISMIRA and LGRA with box-like FCIR versus varying box-size in terms of $\alpha$ for the 4-cell network scenario according to Fig. \ref{fig:sim_Topology3}(b).}
	\label{fig:sim_outage_versus_alpha_4cells2}
	\end{figure}

\begin{figure}
		\centering
		\includegraphics [width=254pt,height=110pt]{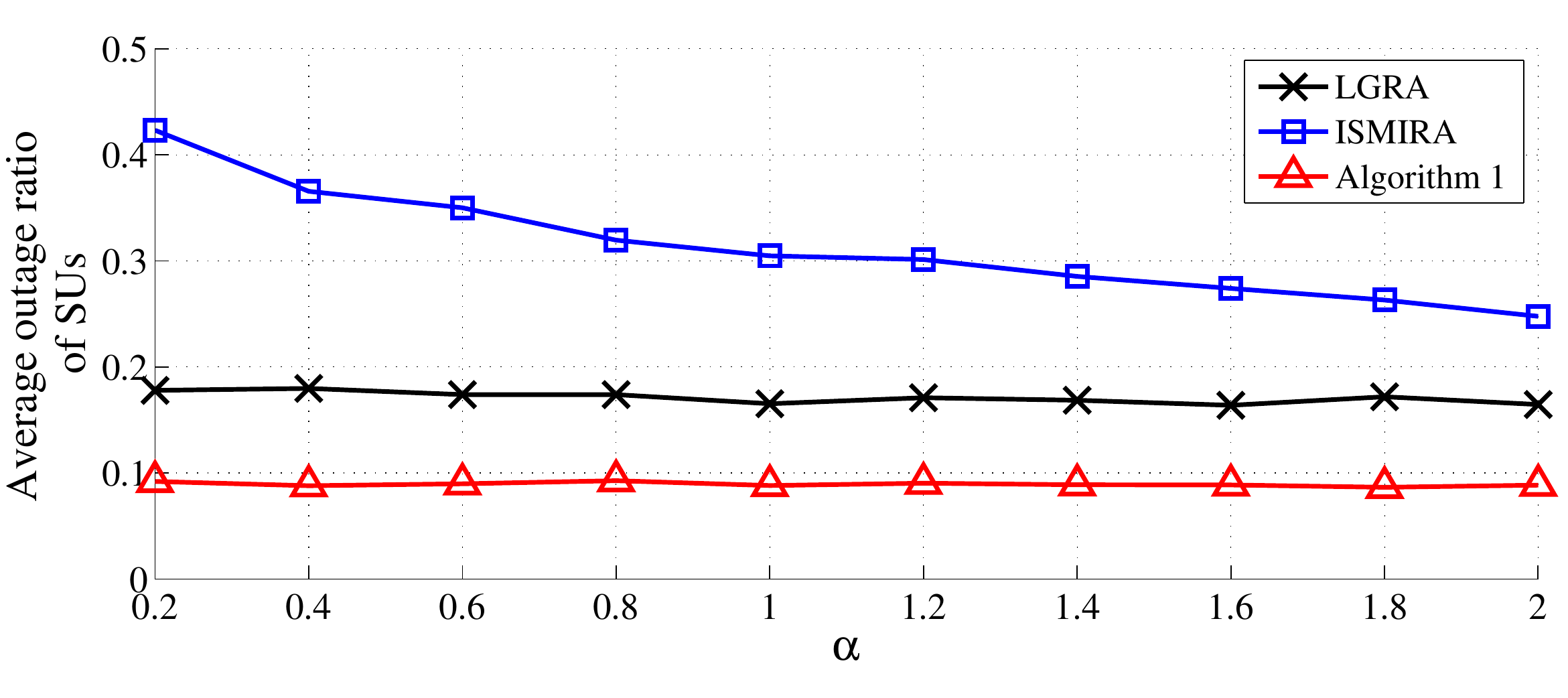}\\ 
		\includegraphics [width=254pt,height=110pt]{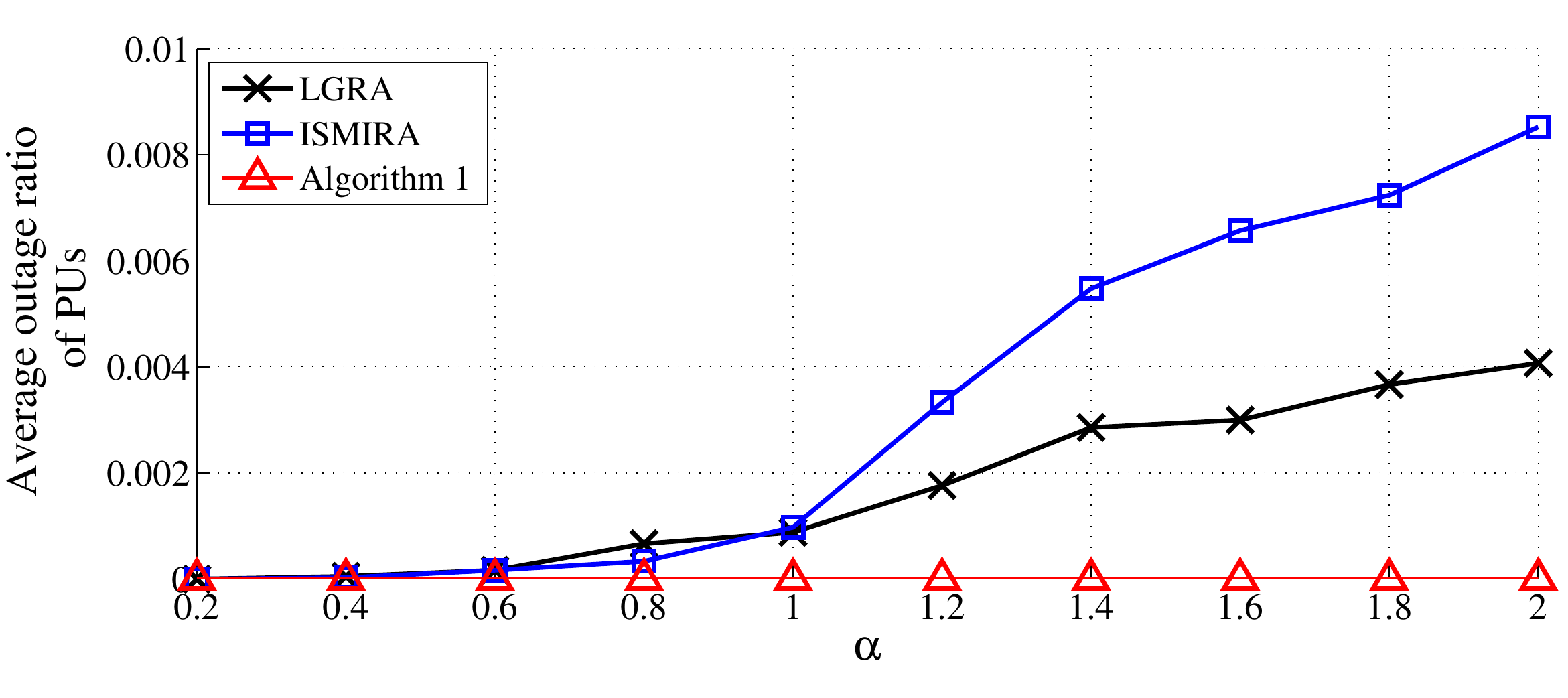}\\ %
		\caption{Average outage ratios of SUs and PUs for Algorithm 1 and for ISMIRA and LGRA with box-like FCIR versus varying box-size in terms of $\alpha$ for the  ad-hoc CRN scenario according to Fig. \ref{fig:sim_Topology3}(c).}
	\label{fig:sim_outage_versus_alpha_adhoc}
	\end{figure}

	To show how the so called ITL values of the PBSs affect the performance of the existing algorithms in comparison to our proposed JPAC algorithm, we consider three different network scenarios. For the first two ones, 20  PUs together with 20  SUs are randomly spread in the coverage area of their corresponding BSs according to the 4-cell network scenarios depicted in Figs. \ref{fig:sim_Topology3}(a) and \ref{fig:sim_Topology3}(b). In the third scenario, 28  PUs together with 28 SUs are randomly located in the ad-hoc network depicted in Fig. \ref{fig:sim_Topology3}(c). Figs. \ref{fig:sim_outage_versus_alpha_4cells1}, \ref{fig:sim_outage_versus_alpha_4cells2}, and \ref{fig:sim_outage_versus_alpha_adhoc} show the average outage ratios of the SUs and PUs for the network scenarios according to Figs. \ref{fig:sim_Topology3}(a), \ref{fig:sim_Topology3}(b), and \ref{fig:sim_Topology3}(c), respectively, for our proposed JPAC algorithm and ISMIRA and LGRA for different values of $\alpha$ ranging from 0.2 to 2 with the step size of 0.2. It is seen that, in all the scenarios, the protection of the PUs is violated for all the algorithms which are  based on the assumption of  box-like FCIR with box sizes corresponding to $\alpha>0.4$. On the other hand,  the average outage ratio of the SUs for our proposed JPAC algorithm is significantly smaller than that of ISMIRA and LGRA for $\alpha\leq 0.4$ in all scenarios. Note that the improved outage performance of SUs in case of ISMIRA and LGRA for large values of $\alpha$, in  comparison with our proposed \textbf{Algorithm 1}, leads to the violation of the  protection of PUs in all scenarios. For the case of the ad-hoc network, it is seen from Fig. \ref{fig:sim_outage_versus_alpha_adhoc} that our proposed JPAC algorithm significantly outperforms ISMIRA and LGRA for all values of $\alpha$.
	

\subsection{Performance Evaluation of the Proposed Algorithm for  Maximizing the Aggregate Throughput for Infeasible Systems}
\begin{figure}
		\centering
		\includegraphics [width=254pt,height=110pt]{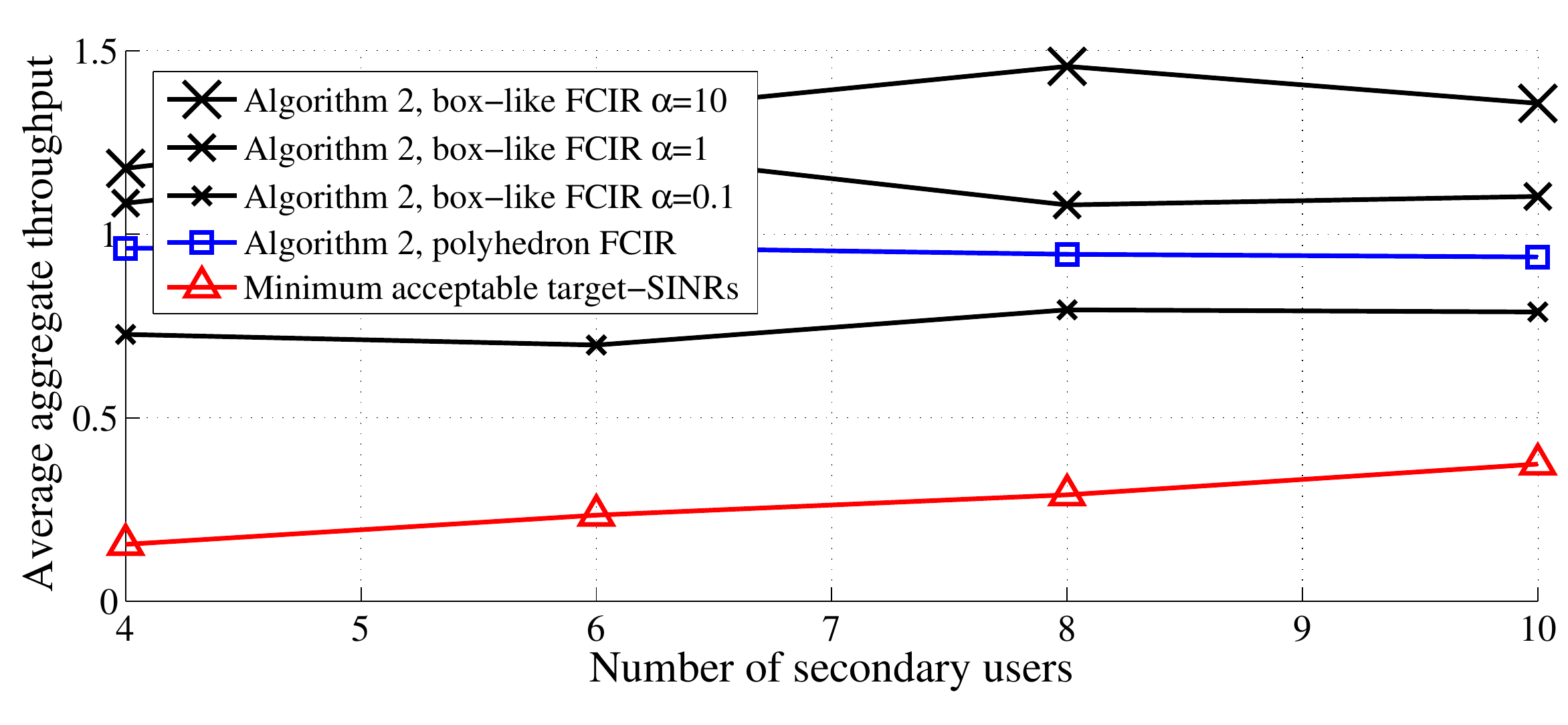}\\ 
		\includegraphics [width=254pt,height=110pt]{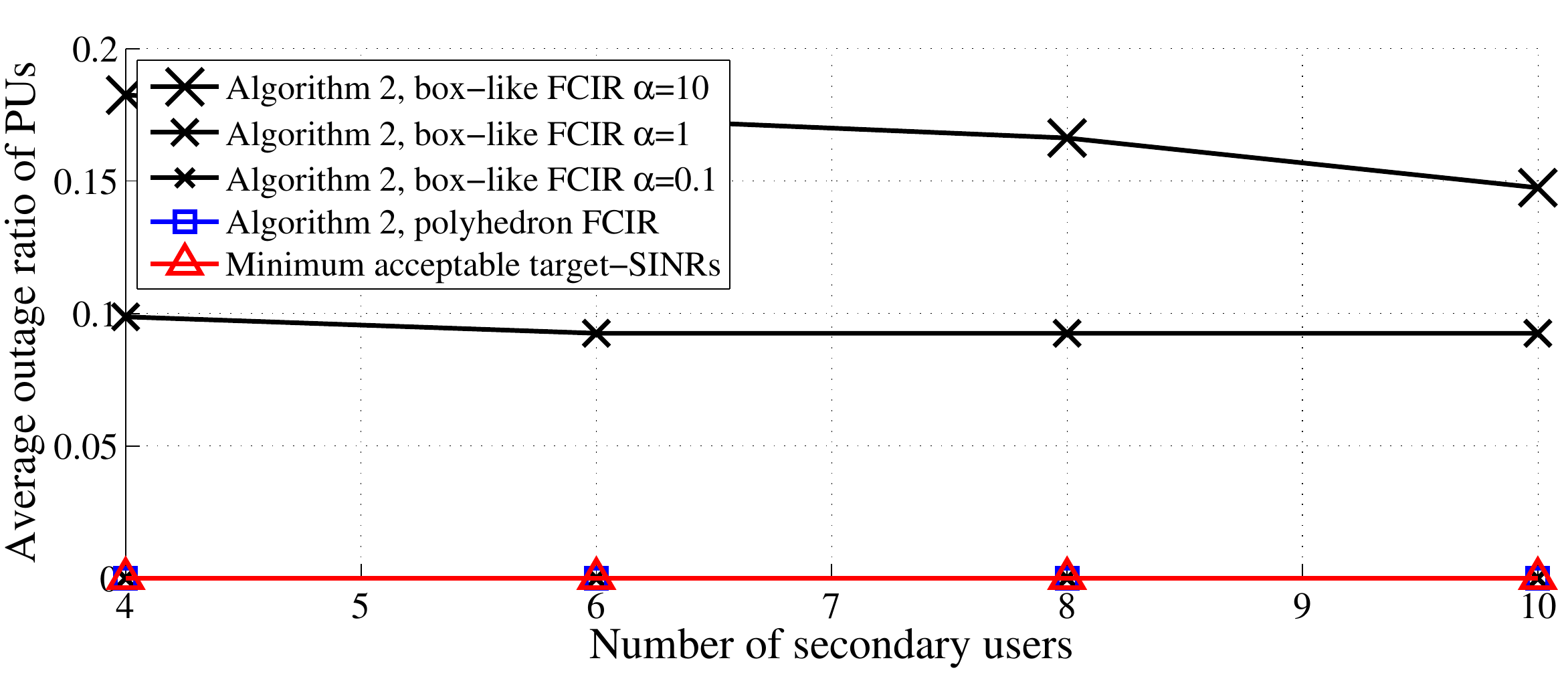}\\ %
		\caption{Average aggregate throughput of SUs and average outage ratio of PUs for Algorithm 2 with the assumption of polyhedron FCIR and box-like FCIRs with $\alpha=0.1$, $\alpha=1$, and $\alpha=10$ and the average aggregate throughput for minimum acceptable target-SINRs for the  4-cell network scenario as shown in Fig. \ref{fig:sim_Topology3}(a).}
	\label{fig:sim_max_throughput_versus_users_1}
	\end{figure}

	To show how the polyhedron and box-like FCIRs affect the performance of power control algorithms for feasible systems, we have considered  8  PUs together with different number of SUs varying from 4 to 10 with the step size of 2 SUs randomly spread in the network according to Fig.  \ref{fig:sim_Topology3}(a) where the target-SINR of each of the PUs and SUs is randomly assigned from the set of $\{-14,-18\}$ dB. We assume $d=150$ m. 
	Under varying number of SUs, Fig. \ref{fig:sim_max_throughput_versus_users_1} shows the average aggregate throughput of the SUs and the average outage ratio of the PUs for feasible systems (where the outage ratio of users is zero when all PUs and SUs assigned with their target-SINRs). This is for \textbf{Algorithm 2} with 
	the assumption of polyhedron FCIR and box-like FCIRs (when the first constraint of \eqref{eq:opt2_feas} is replaced with 	$\sum_{i\in\M^s}{\!\hup{b_m}{i} p_i} \leq \overline{I}^p_m, \forall m\in\B^p$ where $\overline{I}_m^p$ is obtained from \eqref{eq:233}) for different values of $\alpha$ equal to 0.1, 1, and 10. All data are obtained by averaging from 100 independent snapshots. As expected, \textbf{Algorithm 2} with polyhedron FCIR has the best performance as it shows the highest average aggregate throughput subject to the constraint that all PUs remain protected as compared to that of the box-like FCIRs.
	
\section{Conclusion}
	\label{sec:conclusion}

	We have formally defined the feasible region for the cognitive interference caused to the primary receivers by the SUs as an equivalent constraint for the PUs' protection constraint. We have characterized the feasible cognitive interference region (FCIR) and showed that it is generally a polyhedron, as opposed to the box-like FCIR which is commonly assumed in the existing literature to address  the interference management problems in underlay CRNs. This has significant implication on the design of practical interference management schemes for CRNs. 
	The interference management algorithms for underlay CRNs should be designed based on the  polyhedron  region for feasible cognitive interference as characterized in this paper. Based on the obtained value-region of the FCIR, we have devised two power control algorithms for infeasible and feasible CRNs. Through numerical results we have shown that our proposed algorithm significantly outperforms the existing algorithms which are based on the  assumption of the box-like FCIR.
\bibliographystyle{IEEEtran}
\bibliography{Mybib}
\end{document}